\documentclass[
    aps,
    pra,
    reprint,
    superscriptaddress,
    nobibnotes
]{revtex4-2}
\usepackage{amsmath,amssymb}
\usepackage{graphicx}
\usepackage{microtype}
\usepackage[hidelinks]{hyperref}
\usepackage{tikz}

\usetikzlibrary{quantikz2}
\newcommand{\dd}{\mathrm{d}}
\newcommand{\e}{\mathrm{e}}
\newcommand{\calL}{\mathcal{L}}
\newcommand{\norm}[1]{\lVert #1\rVert}
\newcommand{\ip}[2]{\langle #1,#2\rangle}

\providecommand{\Tr}{\mathrm{Tr}}
\usepackage{color,xcolor}
\newcommand{\sun}[1]{\textcolor{black}{#1}}

\hypersetup{colorlinks=true, linkcolor=blue, citecolor=blue, urlcolor=blue}

\makeatletter\def\auto@bib@innerbib{}\makeatother

\begin{document}

\title{Disorder-blind Mpemba acceleration of quantum state preparation}

\author{M Nishanth}
\affiliation{Department of Physics, School of Basic Sciences, Indian Institute of Technology Bhubaneswar, Khurda, Odisha 752051, India}

\author{Shreyas Sadugol}
\affiliation{School of Physical and Chemical Sciences, Queen Mary University of London, London, E1 4NS, United Kingdom}
\affiliation{Massachusetts College of Liberal Arts, North Adams, MA, USA}

\author{Payman Mahmoudi}
\affiliation{School of Physical and Chemical Sciences, Queen Mary University of London, London, E1 4NS, United Kingdom}

\author{Alexander~Yosifov}
\affiliation{Clarendon Laboratory, University of Oxford, Parks Road, Oxford, OX1 3PU, United Kingdom}

\author{Aditya Iyer}
\affiliation{Clarendon Laboratory, University of Oxford, Parks Road, Oxford, OX1 3PU, United Kingdom}

\author{Roopayan Ghosh}
\affiliation{Department of Physics, School of Basic Sciences, Indian Institute of Technology Bhubaneswar, Khurda, Odisha 752051, India}

\author{Jinzhao Sun}
\email{jinzhao.sun.phys@gmail.com}
\affiliation{School of Physical and Chemical Sciences, Queen Mary University of London, London, E1 4NS, United Kingdom}

\date{\today}

\begin{abstract}
Using the strong Mpemba effect to accelerate relaxation by initial-state design requires suppressing the slowest mode, which typically needs prior knowledge of the system. However, in disordered systems, this mode is sample-dependent, so a fixed preparation generally retains a slow-mode overlap. We show that a ``blind'' preparation, designed from the clean chain without knowledge of the disorder realisation or its distribution, can start farther from stationarity yet overtake a reference in typical samples of a dephased disordered XX chain. It requires only the chain geometry and the particle number. At strong dephasing, the density dynamics maps to a resistor chain, and an exact inverse-generator identity links the residual overlap to the spatial resistance profile. We find two distinct thresholds: a finite mean resistance suffices for diffusion, but the residual overlap decays as $L^{-1/2}$ with chain length $L$ only when the resistance variance is finite; rare weak links slow this decay. We predict the fraction of samples in which the blind preparation wins, which is an ensemble-level measure of its reliability. Finite-size calculations support both scaling regimes with interactions at strong dephasing. We give preparation circuits and a density readout for the residual overlap.
\end{abstract}

\maketitle

Preparing many-body quantum states is an important but challenging task in quantum computation. In open-system approaches, a target is encoded as a stationary state, and the system is prepared by relaxing toward it~\cite{verstraete09}, \sun{including dissipative algorithms for thermal and
ground-state preparation~\cite{zhan2026rapid,ding2026simple}}. At late times, relaxation is controlled by the slowest decaying mode, but its amplitude depends on the initial state. A preparation that begins farther from stationarity can thereby overtake one that begins closer---the Mpemba effect \cite{mpemba,luraz,klich,lasanta17,baityjesi,ares23,teza26,summer2026resource}. If its overlap with the slowest mode vanishes, the long-time decay is governed by a faster mode, giving the strong Mpemba effect. This offers a way to shorten equilibration by changing the preparation while keeping the dynamics fixed. The strong effect has been observed in colloidal dynamics \cite{kumar}, a trapped ion \cite{ion,shapira24,joshi2024observing}, and studied in open quantum systems \cite{carollo,review,bagui2026detection}.

Exact mode cancellation generally requires knowledge of the generator or symmetry \cite{carollo,kochsiek22,moroder24,xu,zhou,westhoff2025fast,hong2026geometric} of the quantum system. In the dephased XX chain, for example, symmetry suppresses the slow-mode contribution to Néel-state imbalance in the clean system \cite{gz}. Such state preparation, however, becomes difficult to implement in the presence of disorder, since disorder deforms the slow mode from sample to sample. Previous work has found Mpemba crossings in quasiperiodic systems~\cite{zhou} and disordered spin networks~\cite{shah},  acceleration driven by rare loss-free regions in lossy chains~\cite{longhi}, and \sun{symmetry-restoration Mpemba dynamics robust against on-site disorder on a superconducting processor \cite{xu2026observation}.} On the other hand, hydrodynamic relaxation produces a robust effect in clean chaotic chains \cite{muller}, while a temporary dissipation quench \cite{liuwang} or \sun{coupling to a reset channel~\cite{bao2025accelerating}} can suppress slow modes by changing the dynamics. What remains unknown is whether a fixed preparation designed from the clean system can robustly suppress the slowest relaxation mode across typical disordered realisations, without sample-specific knowledge or modification of the dynamics, and how it scales with system size.

We answer this for a dephased XX chain with random fields or hoppings. We introduce a ``blind'' preparation which is a smooth density modulation orthogonal to the clean slow mode, and compare it with a reference modulation proportional to that mode. At strong dephasing, the density dynamics maps to a one-dimensional (1D) chain of random resistors~\cite{abso,bouchaudgeorges90}: the relaxation rate depends on total resistance, while the scaling of the residual slow-mode overlap depends on its spatial distribution. We find that a finite mean resistance suffices for diffusion, while the typical residual decreases as $L^{-1/2}$ only for finite resistance variance~\cite{agarwal15,gik17}. Rare weak links slow this decrease even when relaxation remains diffusive. Gaussian and stable limit laws give the limiting residual distributions and predict the fraction of samples in which the blind preparation overtakes a reference. Exact quantum calculations support the scaling at finite dephasing, while finite-size calculations with interactions support both regimes at strong dephasing. Circuits that prepare the blind and reference states exactly at fixed particle number, and a density-based readout protocol are also given.

\begin{figure}[t]
\centering
\includegraphics[width=\columnwidth]{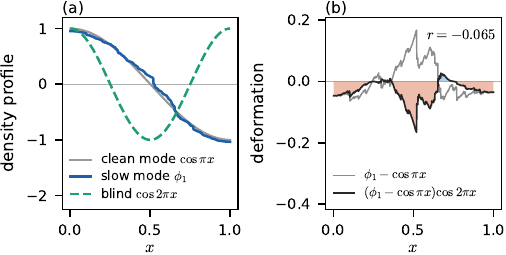}
\caption{(a) In a disordered resistor chain with $L=512$ and $\alpha=1.25$, the slow mode $\phi_1$ is deformed while the blind profile is unchanged; $\alpha$ is the resistance-tail exponent in Eq.~\eqref{eq:tail}. The mode is normalised to unit
$\cos(\pi x)$ coefficient. 
The mode deformation has a small projection onto the blind profile, giving $1/|r|\approx15$ for this sample. (b) Projecting the deformation $\phi_1-\cos\pi x$ onto the blind profile gives $r=-0.065$, normalised by $\int_0^1\cos^2(\pi x)\dd x=1/2$.}
\label{fig:concept}
\end{figure}

\textit{Blind state preparation.---}We consider relaxation under a time-independent Markovian generator $\dot\rho=\mathcal L\rho$ toward a unique stationary state $\rho_{\rm ss}$. Here uniqueness holds within each conserved particle-number sector. Assume a complete biorthogonal mode expansion of $\mathcal L$~\cite{minganti18}
and a simple, isolated slowest nonstationary mode with real $\lambda_1<0$. Its right and left eigenmodes satisfy $\mathcal L R_1=\lambda_1R_1$ and
$\mathcal L^\dagger L_1=\lambda_1L_1$, respectively, with $\langle L_1|R_1\rangle=1$, where $\langle A|B\rangle=\operatorname{Tr}(A^\dagger B)$. A deviation $\delta=\rho(0)-\rho_{\rm ss}$ then evolves at late times as
\begin{equation}
\delta(t)=
\langle L_1|\delta\rangle R_1 e^{\lambda_1t}
+O(e^{\mathrm{Re}\,\lambda_2t}),
\label{eq:modes}
\end{equation}
where $\lambda_2$ is the next-slowest eigenvalue. The amplitude $\langle L_1|\delta\rangle$ controls how much the slow mode contributes to the relaxation. Let $L_1^{(0)}$ denote the left slow mode of the clean reference generator. We call a preparation ``blind'' if its deviation
$\delta_{\rm blind}$ satisfies
$\langle L_1^{(0)}|\delta_{\rm blind}\rangle=0$ and is chosen without information about the disorder realisation or its distribution. Disorder generally changes $L_1$, leaving a nonzero overlap
[Fig.~\ref{fig:concept}(a)]. Relative to a reference preparation $\delta_{\rm ref}$ with the same initial distance from stationarity and nonzero slow-mode overlap, we define the residual
\begin{equation}
r=\frac{\langle L_1|\delta_{\rm blind}\rangle}
{\langle L_1|\delta_{\rm ref}\rangle}
=\frac{\langle L_1-L_1^{(0)}|\delta_{\rm blind}\rangle}
{\langle L_1|\delta_{\rm ref}\rangle},
\label{eq:r}
\end{equation}
where $r$ measures the projection of the slow-mode deformation onto the fixed blind preparation [Fig.~\ref{fig:concept}(b)].

We quantify the preparation's effectiveness using a distance $D$, defined as a norm of the measured deviation from stationarity. Starting from the equally distant deviations in Eq.~\eqref{eq:r}, we scale the blind deviation by $\kappa\ge1$. Provided the slow mode contributes to the measured deviation, their late-time distance ratio approaches $D_{\rm blind}(t)/D_{\rm ref}(t)\to\kappa|r|$.
For $\kappa>1$, a late-time ratio below one guarantees a crossing and the blind preparation is closer to stationarity. This happens when
\begin{equation}
1<\kappa<\mathcal R\equiv 1/{|r|}.
\label{eq:reach}
\end{equation}
We call $\mathcal R$ the reach---the threshold ratio of initial distances. For equal starting distances and $0<|r|<1$, the time advantage approaches $\Delta t=\ln\mathcal R/|\lambda_1|$, corresponding to $\ln\mathcal R$ slow-mode lifetimes~\cite{SM}. If $r=0$, the slow mode is absent and relaxation is governed
by a faster mode. We next calculate how $r$ scales with size and varies across samples.

\textit{Model and the resistance profile.---}We study the dephased disordered XX chain~\cite{gz,esposito05,medvedyeva16,maier19} with $L$ sites, $N=L-1$ bonds, and open boundaries
\begin{equation}
H=\sum_{i=1}^{L-1}J_i(\sigma_i^x\sigma_{i+1}^x+\sigma_i^y\sigma_{i+1}^y)
+\sum_{i=1}^{L}h_i\sigma_i^z,
\end{equation}
where the hoppings $J_i$ or fields $h_i$ are random, and $\sqrt{\gamma}\sigma_i^z$ is local dephasing on each site. After a Jordan-Wigner transformation, the $L\times L$ correlation matrix $M_{ij}=\langle c_j^\dagger c_i\rangle$ obeys an exact closed linear
equation~\cite{znidaric15}, which we use for our calculations (SM Sec.~\ref{sec:Sresistor}). We apply the preceding framework to this closed evolution and measure density relaxation through
$D(t)=[L^{-1}\sum_i\delta n_i(t)^2]^{1/2}$, where
$\delta n_i=n_i-n_{i,\rm ss}$ and $n_i=M_{ii}$. From here on, $\lambda_1$ is the eigenvalue of the slowest mode contributing to $D(t)$. 

Dephasing does not directly change the densities, but damps off-diagonal correlations at rate $4\gamma$. In the diffusive regime, the density-relaxation rates vanish as $L^{-2}$, separating the fast coherence transient from the slow density dynamics. At strong dephasing, eliminating the coherences gives classical nearest-neighbor hopping, with current $\mathcal J_i=G_i(n_i-n_{i+1})$ and conductance~\cite{gz,znidarichorvat}
\begin{equation}
G_i=\frac{8J_i^2\gamma}{4\gamma^2+(h_i-h_{i+1})^2}.
\label{eq:G}
\end{equation}
A bond $i$ is hence a weak link when its hopping is small or the fields on its two sides differ strongly. Its resistance is $\rho_i=1/G_i$: under steady flow, the density drop across bond $i$ is $\mathcal J_i\rho_i$, and resistances add. At finite dephasing, longer-range coherences modify this local resistor. For equal bonds, the density modes are discrete cosines sampled at $x_i=(i-\tfrac12)/L$, and the slowest nonstationary profile is $\phi_1^{(0)}(x)=\cos(\pi x)$. We compare occupation-diagonal states with fixed particle
number $Q$. Their initial deviations from the uniform stationary density $n_{i,\rm ss}=\nu=Q/L$ are $\delta n_i^{\rm ref}(0)=a\cos(\pi x_i)$ and $\delta n_i^{\rm blind}(0)=a\cos(2\pi x_i).$ The reference follows the clean slow mode, while the blind profile
is orthogonal to it by design. Here $a$ is the density amplitude; Appendix B prepares these profiles on a quantum circuit. The two profiles have equal root-mean-square norms; scaling the blind profile by $\kappa$ makes it start $\kappa$ times farther from stationarity. A Néel state has only an $O(1/L)$ normalised overlap with smooth modes (SM Sec.~\ref{sec:Schecks}), so we use smooth density modulations and density-profile readout.

In a disordered sample, calculating $r$ requires the spatial deformation of the slow density mode. Here the bond resistances enter via the normalised cumulative profile
\begin{equation}
u_N(x)=
\frac{\sum_{i=1}^{\lfloor xN\rfloor}\rho_i}
{\sum_{i=1}^{N}\rho_i},
\qquad 0\le x\le 1.
\label{eq:u}
\end{equation}
This is the fraction of the total resistance to the left of $x$. We write the sample-mean
resistance as $\bar\rho_N=N^{-1}\sum_{i=1}^N\rho_i$.
In the continuum limit, a clean chain has $u(x)=x$; disorder produces deviations from this line while preserving $u(0)=0$ and $u(1)=1$. Their magnitude is measured by $\omega=\sup_{0\le x\le1}|u(x)-x|$.

\begin{figure}[t]
\centering
\includegraphics[width=0.9\columnwidth]{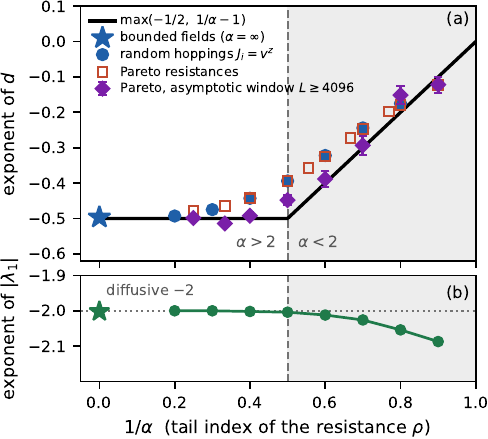}
\caption{(a) Fitted exponent $p_d$ of
$d :=\|\phi_1-\cos(\pi x)\|/\|\cos(\pi x)\|$ versus $1/\alpha$ (i.e., $d \propto L^{p_d}$) for bounded fields $h_0=1$-$16$, random hopping $J_i=v_i^z$, and Pareto resistances. The line gives $\max(-1/2,1/\alpha-1)$, the predicted typical exponent of both $d$ and $|r|$. At $\alpha=2$, there is an additional logarithmic
correction. Diamonds extend Pareto (equivalent to hopping at zero field) to $L=2^{16}$, as heavy tails slow convergence near $\alpha=2$. (b) Fitted exponent $p_\lambda$ of the relaxation rate $|\lambda_1|\propto L^{p_\lambda}$. For random hoppings over $L=64$-$4096$, $|\lambda_1|$ remains close to $L^{-2}$, while $d$ decays more slowly for $\alpha<2$.}
\label{fig:scaling}
\end{figure}

\begin{figure}[t]
\centering
\includegraphics[width=\columnwidth]{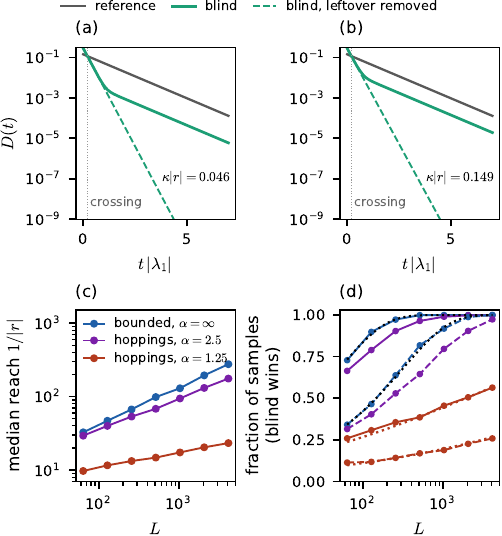}
\caption{Top: Exact quantum dynamics at $\gamma=1$, $L=64$: root-mean-square density distance versus time in slow-mode lifetimes. The
blind preparation starts $\kappa=2$ times farther away, crosses the reference, and approaches the late-time ratio $\kappa|r|$. Dashed curves show the same preparation with its slow-mode component removed. (a) Bounded fields; (b) random hoppings with $\alpha=1.25$. Bottom: Resistor-chain ensembles, with 1500 samples per size. (c) Median reach $1/|r|$. Growth is slower for heavy
tails and the $\alpha=2.5$ data approach slope $1/2$ slowly. (d) Fraction of samples that win for $\kappa=20$ (solid) and $50$ (dashed). Dotted curves are parameter-free predictions:
the Gaussian result Eq.~\eqref{eq:pwin} and the stable-law result Eq.~\eqref{eq:success} with the Appendix C correction for $\alpha=1.25$.}
\label{fig:mpemba}
\end{figure}

\textit{Rate and mode shape.---}For i.i.d. positive resistances with finite mean $\mu_\rho$, the slowest rate of a long chain is
\begin{equation}
\lambda_1\simeq-\frac{\pi^2}{N^2\bar\rho_N},
\qquad
\bar\rho_N\to\mu_\rho.
\label{eq:rate}
\end{equation}
It also applies to the short-range-correlated disorder here \cite{abso,faggionato,derrida83}. Thus $\bar\rho_N$ sets the leading rate; the spatial arrangement enters via relative corrections $O(\omega)$ [Fig.~\ref{fig:scaling}(b)]. For bounded uniform random fields, the leading expression recovers the disorder-averaged gap in \cite{gz}. On large scales, the density deviation evolves as $\partial_t\delta n=N^{-2}K_W\delta n$, with no flux at the ends and
\begin{equation}
K_W=\partial_x\partial_W,
\qquad W(x)=\bar\rho_N u(x),
\end{equation}
where $W(x)=N^{-1}\sum_{i=1}^{\lfloor xN\rfloor}\rho_i$ is the cumulative resistance and $\partial_W$ is differentiation w.r.t. accumulated resistance. For a smooth profile with $W'>0$, it reduces to
$\partial_W=(W')^{-1}\partial_x$. On the lattice, the corresponding relation gives the bondwise density drop (SM Sec.~\ref{sec:Srate}). The eigenfunctions of $K_W$ are the density relaxation modes. Weak links can make $K_W$ far from the clean generator, but its inverse has an exact resistance-weighted form. For real profiles, define $\langle g|f\rangle=\int_0^1g(x)f(x)\dd x$ and $F_f(x)=\int_0^x f(y)\dd y$. For mean-zero $f$ and $g$
\begin{equation}
\langle g|K_W^{-1}f\rangle
=-\int_0^1F_f(x)F_g(x)\,\mathrm{d}W(x).
\label{eq:green}
\end{equation}
This inverse is linear in the resistance profile and the same identity holds bond by bond on the lattice (SM Sec.~\ref{sec:Sprofile}). After removing the overall resistance scale, its difference from the clean inverse is $O(\omega)$, even for very weak links. Since $K_W$ and its inverse share the nonuniform eigenfunctions, this controls the slow-mode deformation. For each disorder realisation, let $\phi_1(x)$ be the slowest nonuniform eigenfunction of $K_W$, normalised so that its $\cos(\pi x)$ is one. Writing $\phi_1=\cos(\pi x)+\sum_{k\ge2}c_k\cos(k\pi x)$, the blind preparation $\cos(2\pi x)$ selects the coefficient $c_2$, so $r=c_2$. For $\omega\ll1$, first-order perturbation theory  gives
\begin{equation}
\begin{aligned}
r&=\int_0^1g(x)\,\mathrm{d}[u(x)-x]+O(\omega^2),\\
g(x)&=\tfrac43\sin(\pi x)\sin(2\pi x).
\end{aligned}
\label{eq:main}
\end{equation}
Eq.~\eqref{eq:main} expresses the residual as a weighted measure of spatial resistance fluctuations, with $g(x)$ setting the sensitivity to their position. The other coefficients $c_k$ are given in SM Sec.~\ref{sec:Sslow}. As an initial example, consider a weak bond at \(x_0\) with large resistance $\rho_{\mathrm w}$ but small resistance share \(s_{\mathrm w}=\rho_{\mathrm w}/\sum_i\rho_i\ll1\), while the rest is uniform. 
Then \(r\simeq s_{\mathrm w}g(x_0)\): its first-order contribution vanishes at the ends and centre and is largest near $x_0\simeq0.30$ or $0.70$, as follows analytically from $g$. Multiplying all resistances by a common factor changes the relaxation time but leaves $r$ unchanged. The residual fixes the late-time distance ratio and reach in each sample, while its distribution gives the success probability across samples. In Fig.~\ref{fig:mpemba}(a,b), a blind state starting twice as far away crosses the reference and then decays with the predicted ratio $2|r|$.

\textit{Resistance tails and residual scaling.---}We now determine how the resistance distribution controls the typical residual as the chain grows. For i.i.d. resistances with $\mu_\rho$, write
$S_m=\sum_{i=1}^m(\rho_i-\mu_\rho)$ for the accumulated excess resistance. The normalised profile obeys
$u(x)-x\simeq[S_{\lfloor xN\rfloor}-xS_N]/(N\mu_\rho)$. Subtracting $xS_N$ pins these fluctuations to zero at both ends.
Their size depends on how frequently resistive bonds occur. For power-law disorder, let
\begin{equation}
\Pr(\rho_i>t)\sim A_\rho t^{-\alpha},
\qquad t\to\infty.
\label{eq:tail}
\end{equation}
Here $A_\rho>0$ is the tail amplitude and $\alpha$ controls the frequency of large resistances. We consider $\alpha>1$, for which the mean is finite. Bounded resistances belong to the finite-variance regime, which we denote $\alpha=\infty$. For $\alpha>2$, many bonds contribute to the profile fluctuations, and the central limit theorem gives a typical residual proportional to $L^{-1/2}$. For $1<\alpha<2$, the mean remains finite but the
variance diverges. The largest of $N$ resistances is of order $N^{1/\alpha}$ and carries a fraction of order $N^{1/\alpha-1}$ of the total resistance. These rare weak links dominate the profile fluctuations, giving $|r|\sim L^{1/\alpha-1}$. At $\alpha=2$, the marginal divergence of the variance gives $|r|\sim\sqrt{\ln L/L}$. These scalings follow by applying the Gaussian and stable
limit laws~\cite{bill,whitt} to Eq.~\eqref{eq:main}
(SM Sec.~\ref{sec:Suneven}). Fig.~\ref{fig:scaling}(a) tests the corresponding slow-mode distortion exponents (Table~\ref{tab:regimes} in Appendix A).

The residual introduces a second threshold in the diffusive regime: a finite mean resistance suffices for diffusion, while finite variance gives the $L^{-1/2}$ improvement of the preparation.
For every $\alpha>1$, $r$ vanishes in probability, so a blind preparation beats a reference with probability tending to one for any fixed starting-distance ratio $\kappa>1$. From the asymptotic exponents, halving the typical residual requires a 32-fold increase in $L$ at $\alpha=1.25$, versus a fourfold increase for finite-variance disorder. The regime can be determined from the microscopic disorder through Eq.~\eqref{eq:G}  (SM Sec.~\ref{sec:Suneven}). Independent fields uniform in $[-h_0,h_0]$ at fixed hopping give bounded resistances. For random hoppings $J_i=v_i^z$, with independent $v_i$ uniform on $(0,1]$ and $h_i=0$, $\alpha=1/(2z)$. For independent exponentially distributed spacings $\ell_i$ of mean $\ell$, tunneling $J_i=J_0e^{-\ell_i/\xi}$ at uniform fields
gives $\alpha=\xi/(2\ell)$.
In this example, relaxation is diffusive for $\xi>2\ell$, whereas the residual follows $L^{-1/2}$ only for $\xi>4\ell$.

\textit{Success probabilities.---}The residual distribution determines how often the blind preparation wins. For an initial distance ratio $\kappa>1$, we define $P_{\rm win}(\kappa)=\Pr(\kappa|r|<1)$ as the fraction of samples in which it has a late-time advantage over the reference. When the resistance variance is finite, $\sqrt Nr$, viewed across disorder realisations, converges to a zero-mean Gaussian with standard deviation $s=2\sigma/(3\mu_\rho)$. Here $\sigma^2$ is the asymptotic variance per bond of the total resistance, including correlations (Appendix A). The Gaussian approximation hence gives
\begin{equation}
P_{\rm win}(\kappa)\simeq
\operatorname{erf}\!\left(
\frac{\sqrt N}{\sqrt{2}\,\kappa s}
\right).
\label{eq:pwin}
\end{equation}
For independent resistances with $1<\alpha<2$, rare weak links instead give a symmetric stable limit~\cite{whitt,st,gnedenko54}
\begin{equation}
N^{1-1/\alpha}\,r
\xrightarrow{\mathrm d}
s_\alpha Z_\alpha,
\label{eq:stable}
\end{equation}
where $Z_\alpha$ is a standard symmetric $\alpha$-stable random variable describing the spread of $r$ over disorder realisations, and the scale $s_\alpha$ depends only on $\alpha$, the tail amplitude $A_\rho$ and the mean $\mu_\rho$ (Appendix A). The limit is symmetric because $g(1-x)=-g(x)$: weak links at mirror positions contribute with opposite signs. Writing $F_\alpha$ for the cumulative distribution function of $Z_\alpha$, the corresponding prediction is
\begin{equation}
P_{\rm win}(\kappa)\simeq
2F_\alpha\!\left(N^{1-1/\alpha}/(\kappa s_\alpha)\right)-1.
\label{eq:success}
\end{equation}
For a specified resistance distribution and ratio $\kappa$, this gives the fraction of samples in which the blind preparation wins. Fig.~\ref{fig:mpemba}(c,d) compares the resulting reach and success probabilities with resistor-chain ensembles.

\textit{Interactions.---}Adding $V\sum_i\sigma_i^z\sigma_{i+1}^z$ of strength $V$ to $H$ breaks the closed evolution of $M$. At strong dephasing, the dynamics becomes a classical lattice gas whose hopping rates depend on the neighboring spins~\cite{poletti12,levi16,Medvedyeva2016}
\begin{equation}
G_i(\sigma)=
\frac{8J_i^2\gamma}
{4\gamma^2+
[h_{i+1}-h_i+V(\sigma_{i+2}^z-\sigma_{i-1}^z)]^2}.
\label{eq:Gint}
\end{equation}
Forward and reverse rates are equal, so the stationary distribution remains uniform within each particle-number sector. The slow mode is now a many-body function, but its density profile still determines $r$ for weak density modulations. In a clean chain, the reflection-even blind preparation is orthogonal to a reflection-odd slow mode; this cancellation holds for all interaction strengths studied here. At half filling, particle-hole symmetry gives $r(V)=r(-V)$ in each sample, so the leading correction near $V=0$ is quadratic (SM Sec. \ref{sec:Sinteract}). For bounded fields and finite $V$, the rates remain proportional to $J_i^2$ up to bounded positive factors, preserving the resistance-tail exponent generated by weak hoppings.

\begin{figure}[t]
\centering
\includegraphics[width=\columnwidth]{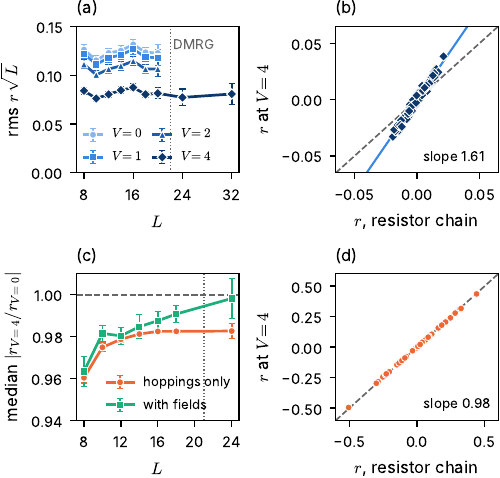}
\caption{Interacting strong-dephasing lattice gas at half filling and $\gamma=4$. Exact diagonalisation is used for $L\le20$; DMRG~\cite{schollwock11,tenpy} is used beyond the dotted line. (a) Bounded fields
$h_i\in[-4,4]$: $\sqrt{L}\,\mathrm{rms}(r)$ is approximately independent of $L$ for the interaction strengths shown, consistent with $|r|\sim L^{-1/2}$. (b) At $V=4$, the same-sample residuals are compared with a resistor chain using configuration-averaged
rates; the solid fit has slope $1.61$ and the dashed line has unit slope. (c) For heavy-tailed hoppings with $\alpha=1.25$, the median $|r(V=4)/r(V=0)|$ is shown with and without fields. (d) At $V=4$ and $L=24$, the heavy-tailed residuals are compared with the averaged-rate chain; the fit has slope $0.98$.
}
\label{fig:interacting}
\end{figure}

\textit{Numerical results.---}We test the scaling predictions using resistor chains up to
$L=2^{16}$. Fig.~\ref{fig:scaling} shows that the size dependence of the slow-mode deformation changes at $\alpha=2$, while the rate retains its diffusive $L^{-2}$ scaling. Exact quantum calculations support this at finite dephasing: for bounded fields at $\gamma=1$, the mean $|r|$ has fitted exponent $-0.49\pm0.02$ (SM Fig. \ref{fig:bounded}), while
heavy-tailed hoppings give a more slowly decreasing slow-mode distortion (SM Fig. \ref{fig:hopping}). Fig.~\ref{fig:mpemba}(a,b) shows the resulting Mpemba effect: a blind preparation starting twice as far from stationarity overtakes the reference, with a late-time distance ratio $2|r|$. Across resistor-chain ensembles, the median reach and success probability follow the predicted size dependence [Fig.~\ref{fig:mpemba}(c,d)]. The heavy-tail comparison includes the finite-size approximation of Appendix C.

The interacting extension in Fig.~\ref{fig:interacting} supports persistence of both residual-scaling regimes at strong dephasing (SM Sec. \ref{sec:Sinteract}). For bounded fields, $\sqrt{L}\,\operatorname{rms}(r)$ is approximately independent of size for the interaction strengths studied: interactions lower the prefactor while the data remain consistent
with $L^{-1/2}$ residual scaling [Fig.~\ref{fig:interacting}(a)].
At $V=\gamma=4$, sample-by-sample residuals track the
configuration-averaged resistor chain [Fig.~\ref{fig:interacting}(b)]. For heavy-tailed hoppings, the estimated median is close to unity at sizes $16\le L\le24$, with and without bounded fields [Fig.~\ref{fig:interacting}(c)]. The averaged-rate prediction tracks these residuals closely [Fig.~\ref{fig:interacting}(d)]. The blind and reference states can be prepared at fixed particle number without sample-specific information, and the residual $r$ can be measured from site-resolved densities (Appendix B).

\textit{Discussion.---}Our results make initial-state design a control resource for relaxation-based many-body state preparation. A profile chosen from the clean system can accelerate equilibration in typical disordered realisations while the generator remains fixed. This offers a route to improve equilibration when disorder varies between experimental runs or when characterising each sample is impractical. Diffusive transport alone is hence insufficient to assess preparation performance: fluctuations in the spatial distribution of resistance also matter. \sun{
Specifically, rare weak links can leave macroscopic diffusion intact while substantially slowing the convergence of disorder-blind mode suppression depending on the heavy tail.} The circuit preparation and density readout of Appendix~B make the predicted overlap suppression testable without reconstructing the slow mode. The predicted success probability
quantifies the resulting reliability across an ensemble. \sun{This addresses a source of uncertainty distinct from state-preparation errors, whose impact on Mpemba acceleration is investigated
\cite{mackinnon2026robustness}. Combining these two sources of uncertainty can provide further assessment of experimental performance in state preparation.} Establishing how broadly this strategy extends to various interacting models, higher dimensions, and weaker dephasing is an interesting next step.  

\textit{Acknowledgements.---}A.Y. and A.I. are supported by UKRI Future Leaders Fellowship (Grant No. 10128920).

\textit{Data availability.---}The code and data supporting the findings of this work are available at \url{https://github.com/NotNIsh7/mpemba_blind_disorder}. 

\bibliography{References}

\newpage

\onecolumngrid

\begin{center}
   \Large\textbf{End Matter}
\end{center}

\twocolumngrid

\textit{Appendix A: Scaling and distribution.---}Table~\ref{tab:regimes} summarizes the typical size dependence of the slowest rate and blind residual. We next give the disorder-dependent scale factors for the Gaussian and stable
limits.

\begin{table}[ht]
\caption{Typical scaling of the slowest rate and the blind residual
for independent resistances with tail as in Eq.~\eqref{eq:tail}.
The residual scalings establish a second threshold at
$\alpha=2$ within the diffusive regime. At $\alpha=2$, the
resistance variance diverges logarithmically. The finite-variance
row also applies to bounded resistances and the short-range
correlations generated by independent random fields.}
\label{tab:regimes}
\begin{ruledtabular}
\begin{tabular}{llll}
Tail & Resistances & Rate $|\lambda_1|$ & Residual $|r|$\\
\hline
$\alpha>2$
& finite variance
& $L^{-2}$
& $L^{-1/2}$\\
$\alpha=2$
& marginal
& $L^{-2}$
& $\sqrt{\ln L/L}$\\
$1<\alpha<2$
& finite mean only
& $L^{-2}$
& $L^{1/\alpha-1}$\\
$\alpha<1$
& infinite mean
& anomalous
& no claim\\
\end{tabular}
\end{ruledtabular}
\end{table}

For stationary resistances with summable covariances, the finite-variance scale is determined by
\[
\begin{aligned}
\sigma^2
&=\lim_{N\to\infty}\frac{1}{N}
\operatorname{Var}\left(\sum_{i=1}^N\rho_i\right)\\
&=\operatorname{Var}(\rho_1)
+2\sum_{m\ge1}\operatorname{Cov}(\rho_1,\rho_{1+m}).
\end{aligned}
\]
For independent resistances this reduces to the single-bond variance. Since the weight in Eq.~\eqref{eq:main} satisfies
$\int_0^1g(x)^2\dd x=4/9$, the Gaussian scale is
$s=2\sigma/(3\mu_\rho)$. For independent resistances with tail \eqref{eq:tail} and $1<\alpha<2$, the stable scale is
\[
s_\alpha
=\frac{\sigma_\alpha}{\mu_\rho}
\left(\int_0^1|g(x)|^\alpha\,\dd x\right)^{1/\alpha},
\
\sigma_\alpha^\alpha
=A_\rho\Gamma(1-\alpha)\cos(\pi\alpha/2).
\]
Here $\Gamma$ is the gamma function, $\sigma_\alpha$ sets the scale of centred resistance sums, whereas $s_\alpha$ sets the dimensionless residual scale.
We write $Z_{\alpha,\beta}$ for a standard $\alpha$-stable random variable with skewness $\beta\in[-1,1]$, unit scale and zero location
\[
\mathbb E\,\e^{i\theta Z_{\alpha,\beta}}
=
\exp\!\left\{
-|\theta|^\alpha
\left[1-i\beta\operatorname{sgn}(\theta)
\tan(\pi\alpha/2)\right]
\right\}.
\]
Thus $Z_\alpha=Z_{\alpha,0}$ in Eq.~\eqref{eq:stable}
is symmetric, whereas the centred resistance sum
$S_N=\sum_i(\rho_i-\mu_\rho)$ in Appendix~C has the maximally right-skewed limit $Z_{\alpha,1}$ after rescaling. The former describes the signed profile-weighted numerator of $r$; the latter describes its fluctuating total-resistance denominator. Detailed derivations are given in SM Sec.~\ref{sec:Suneven}.

\textit{Appendix B: Preparation with a quantum circuit and measurement.---}To turn our disorder-blind initial-state design into an experimentally testable control strategy, we must connect the proposed spatial profiles to both a physical preparation and a measurable signature of slow-mode suppression. Two issues arise on the preparation and measurement sides: (i) a prescribed density profile does not uniquely specify a many-body state, and (ii) the residual slow-mode overlap is defined through a sample-dependent left eigenoperator that is generally unknown.

We address (i) by constructing diagonal states with exactly $Q$ particles, whose preparation weights depend only on the chosen profile, particle number, and modulation amplitude. For (ii), we compare the density responses of the blind and reference preparations in the same disorder realisation. In linear response, once a single slow mode dominates, their ratio eliminates the common decay and readout factors and yields the normalised residual overlap $r$. This provides an experimental test of the proposed relaxation mechanism without prior characterisation of the microscopic disorder or reconstruction of its slow eigenmode. 

\begin{figure}[t]
\centering
\resizebox{\columnwidth}{!}{%
\begin{quantikz}[
row sep={0.6cm,between origins},
column sep=0.12cm,
font=\footnotesize]
\lstick{$s_1$}
& \gate[style={fill=blue!12}]{\theta_1 \,} & \ctrl{1} & & \ctrl{1}
& \ctrl{2} & & \ctrl{1} & \ctrl{3}
& & & \gate[style={fill=orange!18}]{Z^{z_1}} & \\
\lstick{$s_2$}
& \gate[style={fill=blue!12}]{\theta_+} & \targ{}
& \gate[style={fill=blue!12}]{\theta_-} & \targ{}
& & \ctrl{1} & \ctrl{1} &
& \ctrl{2} & & \gate[style={fill=orange!18}]{Z^{z_2}} & \\
\lstick{$s_3$}
& \gate[style={fill=blue!12}]{\pi} & & & &
\gate[style={fill=blue!12}]{\theta_3-\pi}
& \gate[style={fill=blue!12}]{\theta_3-\pi}
& \gate[style={fill=blue!12}]{\pi-2\theta_3} & &
& \ctrl{1} & \gate[style={fill=orange!18}]{Z^{z_3}} & \\
\lstick{$s_4$}
& & & & & & & &
\targ{} & \targ{} & \targ{} & \gate[style={fill=orange!18}]{Z^{z_4}} &
\end{quantikz}%
}
\caption{Four-qubit preparation of the blind state with $Q=2$ and $f=(1,-1,-1,1)/\sqrt2$. All qubits start in $|0\rangle$. Boxes denote $R_y$ rotations, with angles specified in SM Sec.~\ref{sec:Sprep}.
The CNOTs onto $s_4$ set its occupation to the parity of the first three sites. Independent fair bits $z_i$ implement the final phase averaging when their record is discarded.}
\label{fig:circuit}
\end{figure}
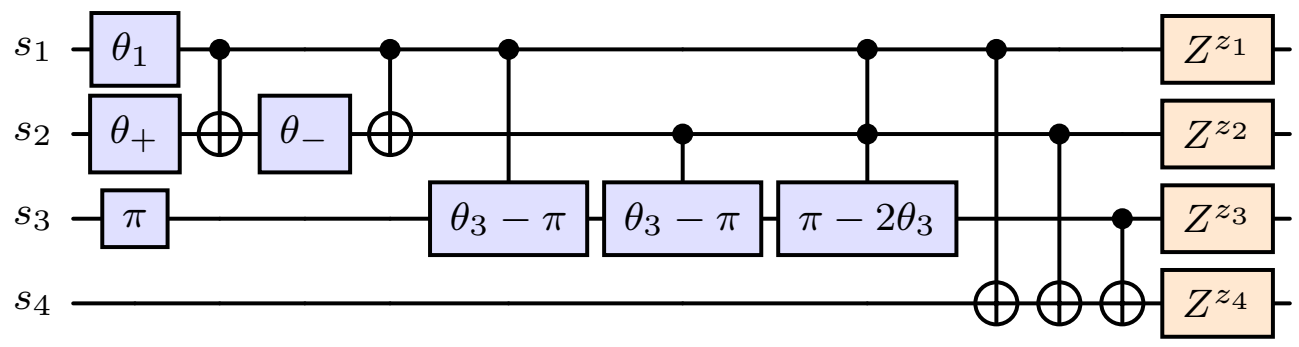

For $f_i=\cos(k\pi x_i)$, with $k=1$ for the reference and $k=2$ for the blind preparation, we prepare a diagonal state with occupation probabilities
\[
p_f(\eta)=\frac{\exp\!\left(\epsilon\sum_i f_i\eta_i\right)}{Z_Q(\epsilon)},\qquad \sum_i\eta_i=Q.
\]
Here $\eta_i\in\{0,1\}$, $Z_Q(\epsilon)$ normalises the distribution, and $\epsilon$ controls the modulation. For small $\epsilon$,
$\langle n_i\rangle-\nu=\epsilon\chi_L f_i+O(\epsilon^2)$, where $\nu=Q/L$ and
$\chi_L=\nu(1-\nu)L/(L-1)$. Thus the density amplitude of the main text is $a=\epsilon\chi_L$ to first order. 
A sequential circuit first prepares $\sum_\eta\sqrt{p_f(\eta)}\,|\eta\rangle$ using rotations controlled by a register that tracks the remaining particle number. Applying independent random $Z$ gates and discarding their classical record removes the coherences, yielding the desired diagonal state exactly. Fig.~\ref{fig:circuit} shows a four-qubit implementation. The circuit fixes $Q$ at its output; an alternative using classical sampling and SWAP gates conserves it throughout (SM Sec.~\ref{sec:Sprep}). Background Hamiltonian evolution and dephasing must be negligible or controlled during the coherent gates.

Having specified how to prepare the initial states, we next describe how to measure their slow-mode suppression.
To measure $r$, prepare each profile at amplitudes $\pm a$ (biases $\pm\epsilon$) in the same disorder realisation and measure $A_1=\sum_i\cos(\pi x_i)(n_i-Q/L)$ from site-resolved densities. Once the slow mode dominates, its common decay and readout factors cancel
\[
\frac{\langle A_1\rangle_{2,+a}(t)-\langle A_1\rangle_{2,-a}(t)}
     {\langle A_1\rangle_{1,+a}(t)-\langle A_1\rangle_{1,-a}(t)}
\xrightarrow[t\to\infty]{}r+O(a^2).
\]
Using both signs of the amplitude cancels the even-order nonlinear response, and the ratio of the two profiles cancels the unknown decay and readout factors; a single sign would leave an $O(a)$ error. The symmetric difference leaves an $O(a^2)$ bias for this state family. The affine preparations in SM Sec.~\ref{sec:Sprep} remove this bias at every allowed amplitude. That section also gives the general circuit construction and explicit rotation angles.

\textit{Appendix C: Finite-size approximation for heavy tails.---}The first-order residual retains the fluctuating total resistance in its denominator
\[
r\simeq
\frac{\sum_{i=1}^N g(i/N)(\rho_i-\mu_\rho)}
     {N\mu_\rho+S_N},
\qquad
S_N=\sum_{i=1}^N(\rho_i-\mu_\rho).
\]
Here $S_N/(\sigma_\alpha N^{1/\alpha})$ converges in distribution
to $Z_{\alpha,1}$. Let $m_\alpha^+$ denote the median of this centred, maximally right-skewed stable variable, $\Pr(Z_{\alpha,1}\le m_\alpha^+)=1/2$. Replacing the denominator by its limiting median estimate gives
\[
P_{\rm win}(\kappa)\simeq
2F_\alpha\!\left(
\frac{N^{1-1/\alpha}
      +(\sigma_\alpha/\mu_\rho)m_\alpha^+}
     {\kappa s_\alpha}
\right)-1.
\]
This approximation accounts for the typical shift of the total resistance without fitted parameters. It omits fluctuations about the median, correlations between numerator and denominator, and higher-order mode deformation. Its accuracy is assessed against the joint stable approximation and resistor-chain
ensembles in SM Fig.~\ref{fig:heavy}.

\clearpage
\onecolumngrid
\begin{center}
{\large\textbf{Supplemental Material for \\
``Disorder-blind Mpemba acceleration of quantum state preparation''}}
\end{center}
\vspace{2mm}
\setcounter{equation}{0}\setcounter{figure}{0}\setcounter{table}{0}\setcounter{section}{0}
\setcounter{secnumdepth}{3}
\renewcommand{\theequation}{S\arabic{equation}}
\renewcommand{\thefigure}{S\arabic{figure}}
\renewcommand{\thetable}{S\arabic{table}}

Here we derive the results of the main text: first the model-independent framework (SM Sec.~\ref{sec:Sblind}), then the reduction of the spin chain to a chain of resistors (SM Sec.~\ref{sec:Sresistor}), the rate (SM Sec.~\ref{sec:Srate}), the resistance profile (SM Sec.~\ref{sec:Sprofile}), the deformed slow mode (SM Sec.~\ref{sec:Sslow}), the statistics of the resistance profile (SM Sec.~\ref{sec:Suneven}), and interacting chains (SM Sec.~\ref{sec:Sinteract}). The preparation and measurement of the blind states are in SM Sec.~\ref{sec:Sprep}. 
The numerical checks can be found in SM Sec.~\ref{sec:Schecks}.

\paragraph{What is known, derived, and observed.}
\label{sec:Sknown}
The rate formula~\eqref{eq:rate} and the limit theorems behind Table~\ref{tab:regimes} are standard~\cite{abso,faggionato,bill,whitt,st}. The profile deviation $u(x)-x$ is the classical corrector of 1D stochastic homogenisation; for finite variance its Brownian-bridge limit is known~\cite{bourgeat,bgmp}, as are Gaussian corrections to spectral problems~\cite{bal08}. Our contributions are the reduction of the disorder-blind Mpemba effect to the residual and its reach, Eqs.~\eqref{eq:r} and \eqref{eq:reach}; the exact linear inverse generator Eq.~\eqref{eq:green}, which controls the slow-mode deformation even when links are arbitrarily weak; its projection onto a fixed blind preparation, Eq.~\eqref{eq:main}; the heavy-tailed regime, whose non-Gaussian corrector separates the threshold $\alpha=2$ for the preparation from $\alpha=1$ for the rate and which we have not found treated in the homogenisation literature; and parameter-free success probabilities in both regimes, Eqs.~\eqref{eq:pwin} and~\eqref{eq:stable}. We also give exact preparation circuits for the blind and reference states, together with a protocol that measures $r$ (SM Sec.~\ref{sec:Sprep}). For interacting chains (SM Sec.~\ref{sec:Sinteract}), the reflection protection and the evenness of $r$ in $V$ at half filling are derived; the persistence of both scaling regimes is a numerical observation for $L\le32$, extended to $L=4096$ for heavy tails through a resistor chain with configuration-averaged rates.

\section{Blind preparation, crossing and reach}
\label{sec:Sblind}

Consider any Markovian dynamics $\dot\rho=\calL\rho$ with a unique stationary state; in our model this holds within each sector of fixed particle number, which the dynamics conserves. We assume that the generator is diagonalisable and has a
simple, isolated slowest nonstationary eigenvalue
$\lambda_1<0$, with right and left modes satisfying
$\calL R_1=\lambda_1R_1$ and
$\calL^\dagger L_1=\lambda_1L_1$, respectively, normalised by $\langle L_1|R_1\rangle=1$. Any deviation from stationarity decays as $\delta(t)=\sum_k\ip{L_k}{\delta}R_k\e^{\lambda_kt}$, so at late times only $\ip{L_1}{\delta}R_1\e^{\lambda_1t}$ survives: the late-time amplitude of a state is its overlap with the left slow mode.

A blind preparation is chosen orthogonal to the slow mode of the clean system $\ip{L_1^{(0)}}{\delta_{\rm blind}}=0$. In a disordered sample the true slow mode differs from $L_1^{(0)}$, and the blind state keeps the overlap
\begin{equation}
r=\frac{\ip{L_1}{\delta_{\rm blind}}}{\ip{L_1}{\delta_{\rm ref}}}=\frac{\ip{L_1-L_1^{(0)}}{\delta_{\rm blind}}}{\ip{L_1}{\delta_{\rm ref}}}.
\end{equation}
Here $\delta_{\rm ref}$ and $\delta_{\rm blind}$ have the same initial measured distance from stationarity. We amplify the blind
deviation by preparing
\[
\rho_{\rm ref}(0)=\rho_{\rm ss}+\delta_{\rm ref},
\qquad
\rho_{\rm blind}^{(\kappa)}(0)
=\rho_{\rm ss}+\kappa\delta_{\rm blind},
\]
with $\kappa\ge1$ chosen so that the states remain positive. 
\sun{ $ \kappa > 1$ is chosen such that crossing indicates that the Mpemba effect takes place, though the Mpemba effect itself does not rely on $\kappa$.
}

Since $D$ is a norm of the measured deviation,
$\kappa=D_{\rm blind}(0)/D_{\rm ref}(0)$.
Explicit positive preparations are given in
Sec.~\ref{sec:Sprep}. When the slow mode contributes to the readout
\begin{equation}
\frac{D_{\rm blind}(t)}{D_{\rm ref}(t)}
\longrightarrow \kappa|r|.
\end{equation}
Thus $1<\kappa<1/|r|$ guarantees a crossing: the blind preparation starts farther away but becomes closer at late times. For $r\ne0$, the late-time distances are
$D_{\rm ref}(t)\simeq A e^{-|\lambda_1|t}$ and
$D_{\rm blind}(t)\simeq\kappa|r|A e^{-|\lambda_1|t}$, with $A>0$. Let $t_{\rm ref}$ and $t_{\rm blind}$ be the times after which
these distances remain below a common tolerance $D_{\rm tol}$. Solving the two exponential decays gives
\begin{equation}
t_{\rm ref}-t_{\rm blind}
\longrightarrow
\frac{\ln[1/(\kappa|r|)]}{|\lambda_1|}
\qquad (D_{\rm tol}\to0).
\end{equation}
The logarithm converts the reduction in slow-mode amplitude into a time advantage. For equal initial distances $\kappa=1$, this
reduces to $\Delta t=\ln(1/|r|)/|\lambda_1|$.
If $r=0$, the blind preparation instead decays through a faster mode, and the two decay rates must be treated separately. In our model the left and right slow density profiles coincide (SM Sec.~\ref{sec:Sresistor}), so we simply write $\phi_1$ for the slow mode, and use the distance $D=[L^{-1}\sum_i\delta n_i^2]^{1/2}$.

\section{From the spin chain to a chain of resistors}
\label{sec:Sresistor}

\paragraph{Exact equation for the densities.} The chain is $H=\sum_iJ_i(\sigma_i^x\sigma_{i+1}^x+\sigma_i^y\sigma_{i+1}^y)+\sum_ih_i\sigma_i^z$ with dephasing $\sqrt\gamma\,\sigma_i^z$ on every site. The Jordan-Wigner transformation turns the hopping term into free fermions $\sigma_i^x\sigma_{i+1}^x+\sigma_i^y\sigma_{i+1}^y=2(c_i^\dagger c_{i+1}+c_{i+1}^\dagger c_i)$, with $\sigma_i^z=2n_i-1$. For this Hermitian jump operator, the real constant shift does not change the dissipator, so dephasing acts as $\sqrt{4\gamma}\,n_i$: it leaves the densities alone and damps every coherence $c_i^\dagger c_j$ ($i\neq j$) at rate $4\gamma$. Since commutators of fermion bilinears are again bilinears, the two-point matrix $M_{ij}=\langle c_j^\dagger c_i\rangle$ obeys a closed equation~\cite{Sznidaric15}
\begin{equation}
\dot M=-i[\mathsf h,M]-4\gamma\,(M-\operatorname{diag}M),\qquad \mathsf h_{ii}=2h_i,\quad \mathsf h_{i,i+1}=\mathsf h_{i+1,i}=2J_i .
\label{eq:closure}
\end{equation}
This is exact and evolves $L^2$ correlation-matrix elements rather than the $4^L$ entries of the full many-body density matrix. Because $\mathsf h$ is real, transposing $M$ maps the generator onto its adjoint, so the left and right slow modes have the same density profile.

\paragraph{Late times: densities only.} The diagonal of Eq.~\eqref{eq:closure} gives a continuity equation, $\dot n_i=\mathcal J_{i-1}-\mathcal J_i$, with the current $\mathcal J_i=4J_i\,{\rm Im}\,M_{i,i+1}$ from site $i$ to $i+1$. The coherence that carries this current obeys
\begin{equation}
\dot M_{i,i+1}=-2i(h_i-h_{i+1})M_{i,i+1}-2iJ_i(n_{i+1}-n_i)-4\gamma M_{i,i+1}+\ldots,
\end{equation}
where the dots couple to coherences between more distant sites. In the diffusive regime, the slow density-relaxation rates vanish as $L^{-2}$ and become small compared with $4\gamma$. A local nearest-neighbor hopping equation follows at strong
dephasing, where couplings to longer-range coherences can be neglected. Setting $\dot M_{i,i+1}=0$ then gives $M_{i,i+1}=-2iJ_i(n_{i+1}-n_i)/[4\gamma+2i(h_i-h_{i+1})]$, and taking the imaginary part
\begin{equation}
\mathcal J_i=G_i\,(n_i-n_{i+1}),\qquad G_i=\frac{8J_i^2\gamma}{4\gamma^2+(h_i-h_{i+1})^2}.
\end{equation}
The densities hop classically $\dot n_i=G_{i-1}(n_{i-1}-n_i)+G_i(n_{i+1}-n_i)$, with no flux through the ends. A bond is a weak link when its hopping is small or when the energies on its two sides differ strongly. These are exactly the bond rates of the effective late-time description used in Ref.~\cite{Sgz} (their Eq.~(15) for the non-interacting chain). At finite $\gamma$, longer-range coherences modify this local approximation. The analytical results below apply to the resistor chain defined by these bond rates; their persistence
at finite dephasing is assessed using the exact correlation matrix dynamics below.

\paragraph{Resistors in series.} Write $\rho_i=1/G_i$. In a steady flow the same current crosses every bond, so the density drops across bond $i$ by $\mathcal J\rho_i$, and the total drop is $\mathcal J\sum_i\rho_i$. Resistances add in series; a single nearly blocked bond can dominate the total.

\section{The relaxation rate}
\label{sec:Srate}

For a uniform chain, the decaying modes are cosines. With $x=(i-\frac12)/L$ one checks directly that $n_i=\cos k\pi x$ solves the hopping equation with no-flux ends, with $\lambda_k=-4G\sin^2(k\pi/2L)\simeq-\pi^2k^2G/L^2$. The slowest is the single arc $\cos\pi x$.

For a disordered chain, label the $N=L-1$ bonds $y_i=i/N$ and define $W(x)=N^{-1}\sum_{y_i\le x}\rho_i$. The lattice equations are, exactly, $\dot n_i=\mathcal J_{i-1}-\mathcal J_i$ and $n_{i+1}-n_i=-\rho_i\mathcal J_i$: the density drops across a bond by current times resistance. Across a large resistance the density therefore jumps, so in a slow mode it is the current that varies smoothly. Since bond $i$ adds $\rho_i/N$ to $W$, the two equations read $\partial_Wn=-N\mathcal J$ and $\partial_tn=-N^{-1}\partial_x\mathcal J$, so that $\dot n=N^{-2}K_Wn$ with
\begin{equation}
K_W=\partial_x\partial_W,\qquad \partial_W=\frac{1}{W'(x)}\,\partial_x,
\end{equation}
and no flux at the ends. A uniform chain with resistance $\bar\rho_N=N^{-1}\sum_i\rho_i$ per bond has $W=\bar\rho_Nx$ and slowest rate $\lambda_1=-\pi^2/(N^2\bar\rho_N)$. Disorder changes this only at relative order $\omega$ (SM Sec.~\ref{sec:Sprofile}), which vanishes for long chains as long as the mean resistance $\mu_\rho$ is finite~\cite{Sabso,Sfaggionato}. The rate thereby depends only on the total resistance
\begin{equation}
\lambda_1\simeq-\frac{\pi^2}{N^2\bar\rho_N},\qquad \bar\rho_N\to\mu_\rho .
\end{equation}
Equivalently, $K_W$ is self-adjoint with $-\ip{f}{K_Wf}=\int_0^1(f')^2/W'\,\dd x$, so $N^2|\lambda_1|$ is the minimum of $-\ip{f}{K_Wf}/\ip{f}{f}$ over mean-zero $f$. On large scales, $1/W'$ is replaced by the harmonic mean of the local conductances $1/\mu_\rho$, which is why only the total resistance enters.

\section{The resistance profile and the inverse generator}
\label{sec:Sprofile}

Define the resistance profile $u(x)=W(x)/W(1)$, which is the fraction of the total resistance to the left of $x$. It is a straight line $u(x)=x$ in a clean chain, and $u(1)=1$ in every sample. Its deviation from a straight line $u(x)-x$ vanishes at both ends, and we denote its largest value $\omega=\sup_x|u(x)-x|$.

\paragraph{Why we work with the inverse generator.} When a few resistances are huge, the generator $K_W$ is not close to the clean one, so ordinary perturbation theory in $K_W$ fails. Its inverse, however, is simple. To solve $K_Wn=f$ for a mean-zero source $f$, integrate once: with no flux at $x=0$, $\partial_Wn=F_f(x)\equiv\int_0^xf$, i.e., $\dd n=F_f\,\dd W$. Physically, the current through each bond is fixed by the source to its left, and the density drops by current times resistance. For a second mean-zero function $g$ with $F_g(x)=\int_0^xg$, an integration by parts then gives
\begin{equation}
\ip{g}{K_W^{-1}f}=-\int_0^1F_f(x)\,F_g(x)\,\dd W(x).
\label{eq:Sgreen}
\end{equation}
The inverse generator is linear in the cumulative resistance. The same identity holds on the lattice, bond by bond $\ip{g}{K^{-1}f}=-\sum_iF_{f,i}F_{g,i}\rho_i$. Writing $W=\bar\rho_Nx+\Delta W$, the change of the inverse generator caused by disorder is $\int_0^1\Delta W\,(fF_g+F_fg)\,\dd x$, which is bounded by $\frac2\pi\bar\rho_N\,\omega\,\norm f\norm g$ (using $\norm{F_f}\le\norm f/\pi$). So it is small whenever the profile is close to a straight line, however large individual resistances are.

\section{The deformed slow mode}
\label{sec:Sslow}

For the uniform chain, the inverse generator has eigenfunctions $\sqrt2\cos k\pi x$ with eigenvalues $\mu_k=-\bar\rho_N/(k\pi)^2$. The slow mode corresponds to the most negative one $\mu_1$, well separated from the rest. Standard first-order perturbation theory~\cite{Skato} gives the deformed slow mode as
\begin{equation}
\phi_1=\cos\pi x+\sum_{k\ge2}c_k\cos k\pi x+O(\omega^2),\qquad c_k=\frac{\ip{e_k}{\delta K^{-1}e_1}}{\mu_1-\mu_k}.
\end{equation}
The numerator follows from Eq.~\eqref{eq:Sgreen}, applied to $W$ and to $\bar\rho_Nx$ and subtracted, with $F_{e_1}=\int_0^xe_1=\sqrt2\sin\pi x/\pi$ and $F_{e_k}=\int_0^xe_k=\sqrt2\sin k\pi x/(k\pi)$, which gives $-\frac{2}{k\pi^2}\int_0^1\sin\pi x\sin k\pi x\,\dd\Delta W$. The denominator is $-\frac{\bar\rho_N}{\pi^2}\frac{k^2-1}{k^2}$. Their ratio, with $\dd\Delta W/\bar\rho_N=\dd[u-x]$, is
\begin{equation}
c_k=\frac{2k}{k^2-1}\int_0^1\sin k\pi x\,\sin\pi x\;\dd\big[u(x)-x\big].
\label{eq:Sck}
\end{equation}

\begin{figure}[htb]
\centering
\includegraphics[width=0.85\textwidth]{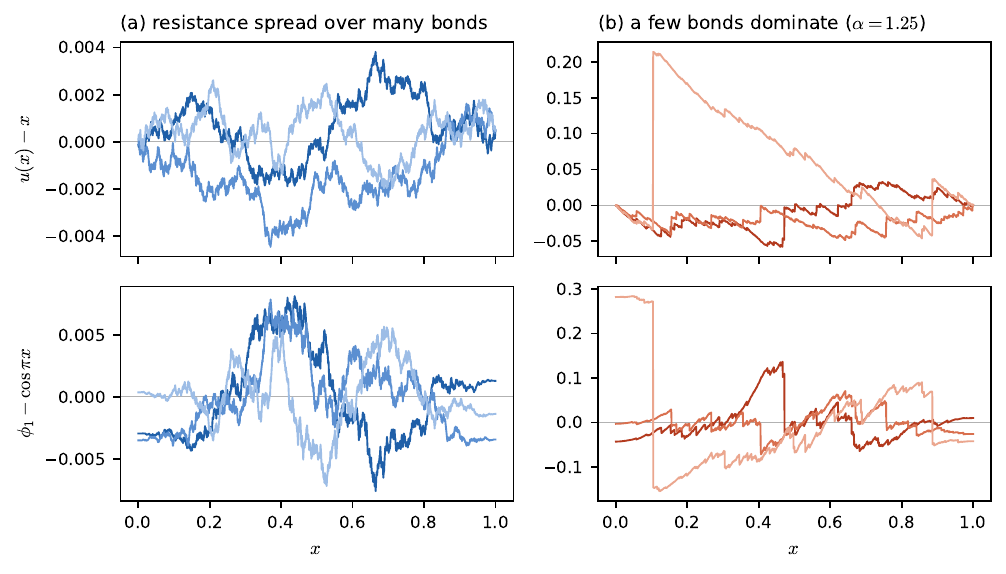}
\caption{What the blind state sees. Top: deviation of the resistance profile from a straight line $u(x)-x$. Bottom: deformation of the slow mode $\phi_1-\cos\pi x$; three samples each, $L=2048$. (a) Bounded disorder: many bonds contribute and the profile wiggles like a random walk pinned at both ends. (b) Heavy-tailed disorder ($\alpha=1.25$): the profile jumps at the weakest links, and the slow mode steps at the same places.}
\label{fig:profiles}
\end{figure}

\paragraph{The residual.} For a blind profile $\cos(2\pi x)$ with the same norm as the reference $\cos(\pi x)$, the residual selects the $k=2$ coefficient. With the normalisation used here, the
reference overlap is fixed, so $r=c_2$. Hence
\begin{equation}
r=\frac43\int_0^1\sin\pi x\,\sin2\pi x\;\dd\big[u(x)-x\big]+O(\omega^2).
\label{eq:Smain}
\end{equation}
Integrating by parts, $r=-\int_0^1 g'(x)[u(x)-x]\,\dd x+O(\omega^2)$ with $g=\frac43\sin\pi x\sin2\pi x$: the residual is a fixed weighted average of the profile deviation. The size of the whole deformation $d=\norm{\phi_1-\cos\pi x}/\norm{\cos\pi x}$ is likewise of order $\omega$. Fig.~\ref{fig:profiles} shows the profile and the deformation in real space.

\paragraph{A single weak link.} Suppose one bond at $x_0$ carries a share $s_w$ of the total resistance and the rest is uniform. Then $u(x)-x=s_w[\Theta(x-x_0)-x]$, so $\dd[u-x]=s_w[\delta(x-x_0)-1]\,\dd x$, and since $\int_0^1g=0$, then
\begin{equation}
r\simeq s_w\,g(x_0)=\tfrac43\,s_w\sin\pi x_0\sin2\pi x_0 .
\label{eq:Swl}
\end{equation}
The weight vanishes at the centre and at both ends and is largest where $\tan^2\pi x_0=2$, i.e., $x_0\simeq0.304$ or $0.696$, where $|g|=16/(9\sqrt3)\simeq1.03$. This is first order in $s_w$ and thus quantitative only for small shares. It shows that in the heavy-tailed regime the residual of a sample is set by the share and position of its few largest resistances.

\begin{figure}[htb]
\centering
\includegraphics[width=0.85\textwidth]{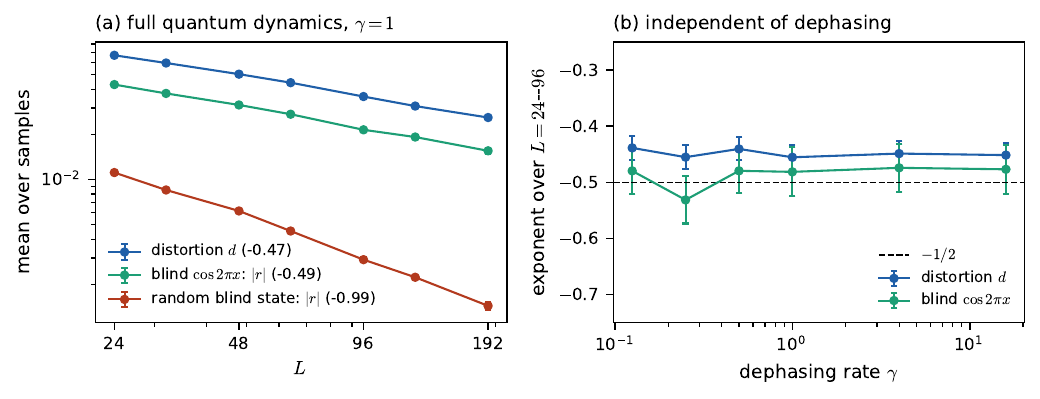}
\caption{Full quantum dynamics, random fields $h_0=2$. (a) Mean slow-mode distortion $d$ and residuals at $\gamma=1$, with fitted exponents, for the blind profile $\cos2\pi x$ and for a random blind state orthogonal to $\cos\pi x$. The latter spreads over all modes and its residual falls as $L^{-1}$ (SM Sec.~\ref{sec:Sslow}). (b) Exponents over $L=24$-$96$ across dephasing rates.}
\label{fig:bounded}
\end{figure}

\paragraph{Other blind profiles.} Any smooth blind profile $\sum_{k\ge2}b_k\cos k\pi x$ with $\sum_kb_k^2=1$ gives $r=\sum_kb_kc_k$ instead. The prefactor $2k/(k^2-1)=\frac43,\frac34,\frac8{15},\dots$ makes $\cos2\pi x$ the most strongly coupled single mode. For finite variance (SM Sec.~\ref{sec:Suneven}), $\sqrt N r$ is Gaussian with standard deviation $s[b]=(\sigma/\mu_\rho)\sqrt{b^{\rm T}\mathsf Mb}$, where $\mathsf M_{kl}=\beta_k\beta_l\int_0^1\sin^2\pi x\sin k\pi x\sin l\pi x\,\dd x$ and $\beta_k=2k/(k^2-1)$. The disorder enters only through $\sigma/\mu_\rho$, so all finite-variance distributions rank blind profiles identically. The largest eigenvalue of $\mathsf M$ gives the least favourable smooth profile $0.975\cos2\pi x-0.221\cos4\pi x$, with $s[b]=0.682\,\sigma/\mu_\rho$, only $2\%$ above $\cos2\pi x$. Rough profiles behave differently: a random blind state orthogonal to $\cos\pi x$, spread over all $\sim N$ lattice modes, catches only a $1/\sqrt N$ share of the deformation, and its residual falls as $L^{-1}$ [Fig.~\ref{fig:bounded}]. The $L^{-1/2}$ statements in the main text therefore refer to smooth blind profiles, which are also the experimentally natural ones.

\section{How uneven is the profile?}
\label{sec:Suneven}

Since $r$ is a fixed weighted average of $u(x)-x$, its size is set by how much the profile fluctuates. Defining $\Delta W(x)=W(x)-xW(1)$ and writing $S_m=\sum_{i\le m}(\rho_i-\mu_\rho)$ for the accumulated excess resistance, the profile deviation is the walk minus its straight-line interpolation
\begin{equation}
\Delta W(x)
=\frac{S_{\lfloor xN\rfloor}-xS_N}{N}+O(N^{-1}).
\label{eq:bridge}
\end{equation}

\begin{figure}[htb]
\centering
\includegraphics[width=0.95\textwidth]{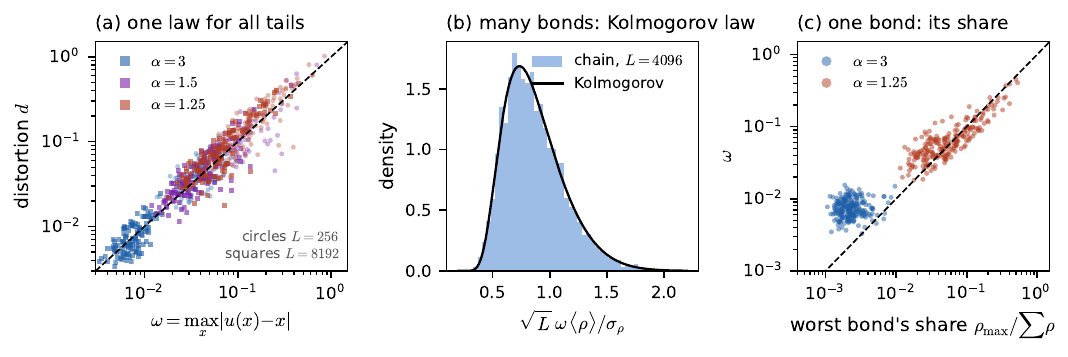}
\caption{The two regimes. (a) The slow-mode distortion $d$ tracks the largest profile deviation $\omega$ sample by sample (random positions with bounded fields, $\xi/\ell=6,3,2.5$, i.e., $\alpha=3,1.5,1.25$; circles $L=256$, squares $L=8192$; line $d=\omega$). (b) For independent resistances uniform in $[1,2]$ at $L=4096$, the rescaled $\omega$ follows Kolmogorov's distribution. (c) At $\alpha=1.25$, $\omega$ is set by the share of the largest resistance; at $\alpha=3$ it is larger, because many bonds contribute.}
\label{fig:ks}
\end{figure}

\paragraph{Finite variance: many bonds.} Suppose the resistances have a finite variance and correlations of finite range, and let $\sigma^2={\rm Var}\,\rho_i+2\sum_{m\ge1}{\rm Cov}(\rho_i,\rho_{i+m})$ be the variance of their sum per bond. Then $S_m$ is a random walk, which on large scales becomes Brownian motion with $S_{xN}\simeq\sigma\sqrt N\,B(x)$~\cite{Sbill}. By Eq.~\eqref{eq:bridge}, $u(x)-x\simeq(\sigma/\mu_\rho\sqrt N)\,B^\circ(x)$, where $B^\circ(x)=B(x)-xB(1)$ is a Brownian bridge (this is the classical 1D homogenisation corrector~\cite{Sbourgeat,Sbgmp}). So $\omega\sim L^{-1/2}$, and the rescaled $\omega$ follows Kolmogorov's distribution [Fig.~\ref{fig:ks}(a,b)]. Inserting into Eq.~\eqref{eq:Smain}
\begin{equation}
\sqrt N\,r\simeq\frac{\sigma}{\mu_\rho}\int_0^1g\,\dd B^\circ=\frac{\sigma}{\mu_\rho}\int_0^1(g-\bar g)\,\dd B,
\end{equation}
a Gaussian with variance $(\sigma/\mu_\rho)^2[\int_0^1g^2-\bar g^2]$, where $\bar g=\int_0^1g$. For $g=\frac43\sin\pi x\sin2\pi x$, $\int_0^1g^2=\frac{16}9\cdot\frac14=\frac49$ and $\bar g=0$, so
\begin{equation}
\sqrt N\,r\xrightarrow{\mathrm d}\mathcal N(0,s^2),\qquad s=\frac23\,\frac{\sigma}{\mu_\rho}.
\end{equation}
Two predictions follow without any free parameters: the median residual is $0.674\,s/\sqrt N$, and a blind state starting $\kappa$ times farther away wins when $|r|<1/\kappa$ (SM Sec.~\ref{sec:Sblind}), which happens with probability
\begin{equation}
P_{\rm win}\simeq{\rm erf}\Big(\frac{\sqrt N}{\sqrt2\,\kappa s}\Big).\label{eq:Spwin}
\end{equation}

\paragraph{Infinite variance: a few weak links.} Now suppose the resistances are independent with a tail $P(\rho>t)\sim t^{-\alpha}$, where $1<\alpha<2$; the mean is finite but the variance is not. Among $N$ bonds the largest resistance is of order $t$ with $N\,P(\rho>t)\sim1$, i.e., $N^{1/\alpha}$, while the total is $\mu_\rho N$. So each of the weakest links carries a share of order $N^{1/\alpha-1}$ of the total resistance, much more than the $N^{-1/2}$ wiggle of many bonds [Fig.~\ref{fig:ks}(c)]. The accumulated excess $S_m$ is then dominated by these jumps and grows as $N^{1/\alpha}$ (the generalised central limit theorem~\cite{Swhitt}), so by Eq.~\eqref{eq:bridge}
\begin{equation}
\omega,\ |r|\sim L^{1/\alpha-1}.
\end{equation}
The profile jumps at the weakest links, and since the density drops across a bond by current times resistance, the slow mode steps at the same places [Fig.~\ref{fig:profiles}(b)]. The residual still vanishes for every $\alpha>1$, so blind preparation works whenever the mean resistance is finite, but reaches its full $L^{-1/2}$ strength only when the variance is finite too. At $\alpha=2$, for a pure $t^{-2}$ tail, the variance of the truncated sum grows as $N\log N$, so $|r|\sim\sqrt{\log L/L}$. For $\alpha<1$ the mean resistance itself diverges, the chain no longer homogenises. This assumes independent resistances; correlated heavy-tailed disorder is tested numerically.

\paragraph{Infinite variance: the distribution of the residual.} Let the resistances be independent with $P(\rho_i>t)\simeq A_\rho t^{-\alpha}$, where $1<\alpha<2$. On the lattice, Eq.~\eqref{eq:Smain} becomes a ratio of two sums. The profile $u$ jumps by $\rho_i/\sum_j\rho_j$ across bond $i$, so $\int_0^1g\,\dd u=\sum_ig_i\rho_i/\sum_i\rho_i$, while $\int_0^1g\,\dd x$ becomes $N^{-1}\sum_ig_i$. Hence
\begin{equation}
r=\frac{\sum_ig_i\rho_i}{\sum_i\rho_i}-\frac1N\sum_ig_i,\qquad g_i=g(x_i).
\end{equation}
The last term is $O(N^{-2})$, because $\int_0^1g=0$ and $g$ vanishes at both ends, so that $\sum_ig_i=O(N^{-1})$. To expose the fluctuations, write each resistance as its mean plus an excess $\rho_i=\mu_\rho+(\rho_i-\mu_\rho)$, and define
\begin{equation}
T_N=\sum_ig_i(\rho_i-\mu_\rho),\qquad S_N=\sum_i(\rho_i-\mu_\rho).
\end{equation}
$T_N$ is the excess resistance weighted by the blind kernel, and $S_N$ is the excess of the total resistance. Then $\sum_ig_i\rho_i=T_N+\mu_\rho\sum_ig_i$ and $\sum_i\rho_i=N\mu_\rho+S_N$. Dropping terms smaller by a factor $N^{-1}$
\begin{equation}
r\simeq\frac{T_N}{N\mu_\rho+S_N}=\frac{T_N/N}{\mu_\rho+S_N/N}.
\end{equation}
For heavy tails both sums are dominated by the largest resistances and grow as $N^{1/\alpha}$ rather than $N^{1/2}$. Writing each on this natural scale $T_N/N=N^{1/\alpha-1}\,(T_N/N^{1/\alpha})$ and likewise for $S_N$, gives
\begin{equation}
r\simeq\frac{N^{1/\alpha-1}\,T_N/N^{1/\alpha}}{\mu_\rho+N^{1/\alpha-1}\,S_N/N^{1/\alpha}} .
\label{eq:Sratio}
\end{equation}
Here $T_N/N^{1/\alpha}$ and $S_N/N^{1/\alpha}$ are random variables of order one, whose stable limits are derived next. The prefactor $N^{1/\alpha-1}$ sets the size of $r$. Because the same prefactor multiplies $S_N$, the denominator fluctuates at the same relative order as $r$ itself; this is the origin of the finite-size correction below. By the generalised central limit theorem for weighted sums~\cite{Swhitt,Sst}, $(T_N,S_N)/N^{1/\alpha}$ converges jointly to $(\int_0^1g\,\dd Z,\,Z(1))$, where $Z$ is a totally skewed $\alpha$-stable L\'evy process whose increments over unit length have scale $\sigma_\alpha=[A_\rho\Gamma(1-\alpha)\cos(\pi\alpha/2)]^{1/\alpha}$ (positive, since both factors are negative). For deterministic $g$, $\int_0^1g\,\dd Z$ is $\alpha$-stable with scale $\sigma_\alpha\norm g_\alpha$, where $\norm g_\alpha=(\int_0^1|g|^\alpha\dd x)^{1/\alpha}$, and skewness $\beta_g=\int_0^1|g|^\alpha{\rm sgn}\,g\,\dd x/\norm g_\alpha^\alpha$~\cite{Sst}. Since $g(1-x)=-g(x)$, $\beta_g=0$ exactly: although every $\rho_i$ is strongly skewed, the residual is symmetric. To leading order the denominator of Eq.~\eqref{eq:Sratio} is $\mu_\rho$, so
\begin{equation}
N^{1-1/\alpha}\,r\;\xrightarrow{\mathrm d}\;s_\alpha Z_\alpha,\qquad s_\alpha=\frac{\sigma_\alpha\norm g_\alpha}{\mu_\rho},
\label{eq:Sstable}
\end{equation}
with $Z_\alpha$ standard symmetric $\alpha$-stable, $\mathbb E\,\e^{i\theta Z_\alpha}=\e^{-|\theta|^\alpha}$. With $F_\alpha$ its distribution function and $m_\alpha$ the median of $|Z_\alpha|$, a blind state starting $\kappa$ times farther away wins with probability $P_{\rm win}\simeq2F_\alpha(z_\kappa)-1$, where $z_\kappa=N^{1-1/\alpha}/(\kappa s_\alpha)$, and the median reach is asymptotically $z_1/m_\alpha$. For random hoppings, $\rho_i=(\gamma/2)v_i^{-2z}$ is exactly Pareto distributed, with $A_\rho=(\gamma/2)^\alpha$ and $\mu_\rho=(\gamma/2)\alpha/(\alpha-1)$, so $s_\alpha$ depends on $\alpha$ alone. For $\alpha=1.25$, $\sigma_\alpha/\mu_\rho=0.331$, $\norm g_\alpha=0.597$, $s_\alpha=0.198$, and $m_\alpha=0.979$.

\begin{figure}[htb]
\centering
\includegraphics[width=0.95\textwidth]{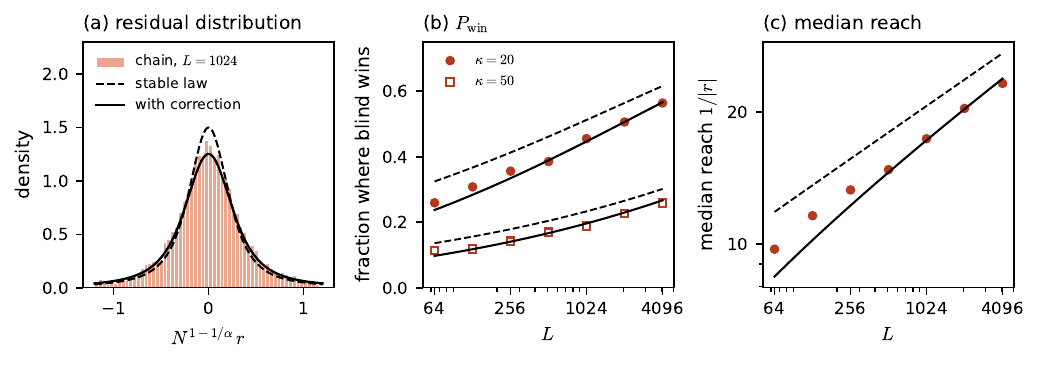}
\caption{Stable-law statistics of the residual for heavy-tailed random hoppings $\alpha=1.25$, compared with exact slow modes of the resistor chain. (a) Distribution of $N^{1-1/\alpha}r$ at $L=1024$ (6000 samples) against the symmetric stable law~\eqref{eq:Sstable} (dashed) and with the finite-size correction~\eqref{eq:Szk} (solid). (b) Fraction of samples in which a blind state starting $\kappa=20$ (filled) or $50$ (open) times farther away wins, from the 1500 samples per size of Fig.~\ref{fig:mpemba}. (c) Median reach, same data. No parameter is fitted.}
\label{fig:heavy}
\end{figure}

\emph{Finite-size approximation.} Eq.~\eqref{eq:Sstable} is approached slowly. The relative fluctuation of the denominator in Eq.~\eqref{eq:Sratio} is of order $N^{1/\alpha-1}$, the same order as $r$, and it is totally skewed: in a typical sample the total resistance lies below $N\mu_\rho$, because the mean is carried by rare weak links, so typical residuals are larger than the leading order suggests. Replacing $S_N/N^{1/\alpha}$ by its median $\sigma_\alpha m_\alpha^+$, where $m_\alpha^+$ is the median of the standard totally skewed stable law ($m_{1.25}^+=-1.996$), gives
\begin{equation}
z_\kappa=\frac{\mu_\rho N^{1-1/\alpha}+\sigma_\alpha m_\alpha^+}{\kappa\,\sigma_\alpha\norm g_\alpha}.
\label{eq:Szk}
\end{equation}
For the cases tested, sampling the joint stable law of the numerator and denominator changes $P_{\rm win}$ by at most $0.01$ relative to the median approximation. This checks the treatment of the fluctuating denominator within the first-order residual formula. Fig.~\ref{fig:heavy} compares both forms with exact slow modes of the resistor chain (the 1500 samples per size of Fig.~\ref{fig:mpemba}): with the correction, $P_{\rm win}$ agrees within $0.022$ ($\kappa=20$) and $0.008$ ($\kappa=50$) for $256\le L\le4096$, and the median reach within $9\%$ at $L=256$ and 3\% for $512\le L\le4096$. The leading order alone is $0.02$-$0.08$ too high over $64\le L\le4096$.

\textit{Rare failures.}
The stable limit describes the distribution of typical rescaled residuals. At fixed $\kappa$, failure probes increasingly remote tails of this distribution as the chain grows, so convergence in distribution alone does not determine the asymptotic failure probability. Samples containing a weak link with an appreciable share of the total resistance can also require higher-order mode deformation. We therefore assess the finite-size success predictions against the resistor-chain ensembles in Fig.~\ref{fig:heavy}.

\paragraph{Where the tails come from.} Random fields $h_i$ uniform in $[-h_0,h_0]$ with $J_i=J$ give $\rho_i=[4\gamma^2+D_i^2]/(8J^2\gamma)$, where $D_i=h_i-h_{i+1}$: bounded resistances, so $\alpha=\infty$. Neighbouring resistances share a field, so the covariance term in $\sigma^2$ matters. Using $\mathbb Eh^2=h_0^2/3$ and $\mathbb Eh^4=h_0^4/5$, one finds ${\rm Var}\,D^2=\frac{28}{45}h_0^4$ and ${\rm Cov}(D_i^2,D_{i+1}^2)=\frac4{45}h_0^4$, so that $\sigma^2=\frac{36}{45}h_0^4/(8J^2\gamma)^2$. For $h_0=2$, $\gamma=J=1$ this gives $\mu_\rho=5/6$, $\sigma^2=0.2$ and $s=0.358$. Random hoppings $J_i=v_i^z$, with $v_i$ uniform on $(0,1]$ and $h_i=0$, give $\rho_i\propto v_i^{-2z}$, hence $P(\rho>t)\propto t^{-1/2z}$ and $\alpha=1/2z$. For independent exponentially distributed spacings $\ell_i$ of mean $\ell$, tunnelling $J_i=J_0e^{-\ell_i/\xi}$ at uniform fields gives
$\rho_i=\rho_0e^{2\ell_i/\xi}$, where $\rho_0=\gamma/(2J_0^2)$. Thus $\Pr(\rho_i>t)=(\rho_0/t)^\alpha$ for $t\ge\rho_0$,
with $\alpha=\xi/(2\ell)$. Independent bounded random fields multiply these resistances
by a bounded positive factor and preserve the tail exponent, but introduce correlations between neighboring bonds. The distributional results above are derived for independent heavy-tailed resistances.

\section{Interacting chains}
\label{sec:Sinteract}

\paragraph{Effective dynamics.} Add $V\sum_i\sigma_i^z\sigma_{i+1}^z$ to $H$. The density equation no longer closes, but the separation of timescales of Sec.~\ref{sec:Sresistor} still applies to the many-body populations. A hop across bond $i$ connects two configurations that differ on two sites, so the coherence between them decays at rate $4\gamma$ and oscillates at their energy difference $\Delta E_i$, which now includes the change of the interaction energy with the two outer neighbours. Eliminating this coherence at strong dephasing gives a classical exclusion process with the rate of Eq.~\eqref{eq:Gint}, $G_i=32J_i^2\gamma/(16\gamma^2+\Delta E_i^2)$, where $\Delta E_i=2(h_{i+1}-h_i)+2V(\sigma^z_{i+2}-\sigma^z_{i-1})$ and $\sigma^z\equiv0$ beyond the ends. The rate depends on $\Delta E_i^2$ and is therefore the same for a hop and its reverse. The generator is then symmetric, and the uniform distribution $\pi$ on configurations with fixed particle number is stationary for every $V$ and every sample. This is the infinite-temperature state of the dephased chain, and it carries no information about the disorder. At $V=0$ the rates do not depend on the configuration and the densities obey the resistor chain of Sec.~\ref{sec:Sresistor}.

\paragraph{The residual.} The left slow mode is a function $f_1(\eta)$ (the left eigenvector of the lattice-gas generator with the slowest nonzero rate; with interactions the density equation does not close, so the slow mode is a function of the whole configuration rather than a density profile) on configurations. A density modulation $b_i$ prepared in local equilibrium, $P=\pi\,[1+\epsilon\sum_ib_i(\eta_i-\nu)]$ with filling $\nu$, has slow-mode amplitude $\epsilon\sum_ib_im_i$ with $m_i=\mathbb E_\pi[f_1(\eta)\,\eta_i]$. The density profile $m_i$ of the slow mode therefore plays the role of $\phi_1$, and
\begin{equation}
r=\frac{\sum_i\cos(2\pi x_i)\,m_i}{\sum_i\cos(\pi x_i)\,m_i}.
\end{equation}
The overlap is $\sum_\eta f_1(\eta)P(\eta)=\epsilon\sum_ib_i\,\mathbb E_\pi[f_1(\eta_i-\nu)]=\epsilon\sum_ib_im_i$, because $\mathbb E_\pi f_1=0$ (the slow mode is orthogonal to the stationary state). The residual is the ratio of these amplitudes for $b_i=\cos(2\pi x_i)$ and $\cos(\pi x_i)$ at equal $\epsilon$. Such a state is the linear-response member of the maximum-entropy family $P\propto\pi\,\e^{\sum_ib_i\eta_i}$ at fixed particle number, which can also be prepared exactly by quantum circuits. SM Sec.~\ref{sec:Sprep} gives these circuits, and its affine states make the linear-response form exact.

\paragraph{Symmetries and reflection.} If the generator and $\pi$ are invariant under $i\to L+1-i$ and $f_1$ is odd, then $\mathbb E_\pi[f_1\,P]=0$ for every reflection-even $P$, at any amplitude and any $V$. In clean chains we find $|r|<10^{-13}$ for all $V$ studied. \emph{Interaction sign.} At half filling, particle-hole conjugation $\eta_i\to1-\eta_i$ reverses every $\sigma^z$ and every hop, so it maps the rates at $V$ onto those at $-V$. It also reverses every zero-sum density deviation, and with it the slow-mode profile, $m_i\to-m_i$. Hence $r(V)=r(-V)$ in every sample, which we confirm to $10^{-14}$ for $L=10$-$14$. The same holds for the full Lindblad generator at any $\gamma$: the map $O\mapsto SC\,O^{\rm T}CS$, with $C=\prod_iX_i$ and $S=\prod_{i\ {\rm odd}}Z_i$, sends the generator at $V$ to the one at $-V$ for real hoppings, and odd density deviations to minus themselves.
The leading interaction correction is therefore of order $V^2$. Consistently, the slope of $r$ against the averaged-rate chain below exceeds one by $0.06$ at $V=1$ and $0.25$ at $V=2$.

\paragraph{Methods.} For $L\le20$ we diagonalise the generator in the half-filled sector exactly (sparse Lanczos; $184\,756$ configurations at $L=20$). For larger $L$ we write minus the generator as a matrix-product operator with four-site terms. It is real symmetric, so the slowest mode is the first excited state of a Hermitian problem. We obtain it with two-site DMRG~\cite{Stenpy}, orthogonalising against the uniform ground state and shifting the operator by a constant so that the excited state has negative energy. The ground state is a fixed-particle-number superposition of all configurations, with entanglement growing only logarithmically, and bond dimension $64$ suffices: at $L=24$ raising it to $96$ changes $r$ in the sixth digit. DMRG reproduces exact diagonalisation to all printed digits for $L=10$-$16$, including samples with a link of $J_i^2=10^{-4}$. Unless stated otherwise $J=1$, $\gamma=4$.

\paragraph{Results.} Table~\ref{tab:interact} collects the residual statistics. For bounded fields, $h_i\in[-4,4]$, the rms of $\sqrt L\,r$ is independent of $L$ for every $V$. Interactions lower it from $0.12$ to $0.08$ at $V=4$ while leaving the exponent at $-1/2$. Sample by sample, $r$ is $98$-$100\%$ correlated with the residual of a resistor chain whose conductances are the configuration-averaged rates, $\bar G_i=\mathbb E_\pi[G_i\,\mathbf 1_{\eta_i\neq\eta_{i+1}}]$, which differ from bond to bond because the interaction energy adds to the local field difference. The ratio between the two settles at $1.6$ for $L=24$-$32$. This enhancement comes from correlations between neighbouring occupations, which the averaged rates neglect. For heavy-tailed hoppings, $J_i=v_i^{0.4}$ ($\alpha=1.25$), interactions change $r$ by less than $2\%$ for $L\ge16$, with and without fields, and $r$ is $99.7$--$100\%$ correlated with the averaged-rate chain. In this regime the residual is dominated by the few weakest links, and interactions keep each rate between $8J_i^2\gamma/[4\gamma^2+(|h_{i+1}-h_i|+2V)^2]$ and $2J_i^2/\gamma$. The resistance of a weak link therefore still scales as $J_i^{-2}$, and its tail exponent is unchanged. The averaged-rate chain can be evaluated up to $L=4096$: over $L=64$-$4096$ its median residual has exponent $-0.230$ at $V=0$ and at $V=4$ ($-0.233$ and $-0.229$ with fields) for $\alpha=1.25$, and $-0.446$ for $\alpha=2.5$. These are the same slowly converging exponents as for non-interacting resistor chains (Table~\ref{tab:checks} predicted $-0.2$ and $-0.5$). The threshold at $\alpha=2$ thus survives interactions.

\begin{table}[htb]
\caption{Interacting chains. Bounded fields: rms of $\sqrt L\,r$ (300 samples for $L\le18$, 100 at $L=20$, 40 and 30 DMRG samples at $L=24$, $32$; relative error about $4\%$, $7\%$, $11\%$, $13\%$). Heavy tails: median of $|r(V=4)/r(V=0)|$ over the same samples (400 for $L\le18$, 40 at $L=24$).}
\label{tab:interact}
\begin{ruledtabular}
\begin{tabular}{lccccccccc}
$L$ & 8 & 10 & 12 & 14 & 16 & 18 & 20 & 24 & 32\\
\hline
bounded, $V=0$ & 0.126 & 0.115 & 0.123 & 0.125 & 0.132 & 0.123 & 0.123 & & \\
bounded, $V=1$ & 0.122 & 0.111 & 0.118 & 0.121 & 0.127 & 0.118 & 0.118 & & \\
bounded, $V=2$ & 0.111 & 0.100 & 0.106 & 0.110 & 0.114 & 0.106 & 0.106 & & \\
bounded, $V=4$ & 0.084 & 0.076 & 0.080 & 0.085 & 0.087 & 0.080 & 0.081 & 0.077 & 0.081\\
heavy, no fields & 0.960 & 0.975 & 0.979 & 0.981 & 0.982 & 0.983 & & 0.983 & \\
heavy, with fields & 0.963 & 0.982 & 0.980 & 0.985 & 0.988 & 0.991 & & 0.998 & \\
\end{tabular}
\end{ruledtabular}
\end{table}

\section{Preparing the blind states on a quantum circuit and measuring them}
\label{sec:Sprep}

Appendix B summarises the preparation and readout protocols and shows the four-qubit circuit in Fig.~\ref{fig:circuit}. Here we give the state definitions, general circuit construction, and
derivations. The reference and blind profiles are
$f_i^{(k)}=\cos(k\pi x_i)$, with $k=1$ and $k=2$, respectively. Both preparations require only the profile, the particle number $Q$, and the modulation amplitude. We consider $0<Q<L$ and write
$P_Q$ for the projector onto the $Q$-particle sector, $d_Q=\binom{L}{Q}$, $\nu=Q/L$, and $\rho_*=P_Q/d_Q$. For a profile satisfying $\sum_i f_i=0$, define
\begin{equation}
A_f=\sum_i f_i(n_i-\nu).
\end{equation}

\paragraph{Two families of states.}
We use two preparation families for complementary purposes. The exponential family is positive for any finite bias and admits
the sequential circuit below; its density modulation follows the chosen profile in linear response. The affine family has a restricted amplitude range but gives exactly linear density and slow-mode responses, allowing the residual to be measured without an amplitude extrapolation.

The exponential family consists of the diagonal states
\begin{equation}
\begin{aligned}
\rho_f(\epsilon)
&=\frac{P_Q e^{\epsilon A_f}P_Q}{Z_Q(\epsilon)}
=\sum_{|\eta|=Q}p_f(\eta)|\eta\rangle\langle\eta|,\\
p_f(\eta)
&=\frac{1}{Z_Q(\epsilon)}
  \prod_i w_i^{\eta_i},
\qquad
w_i=e^{\epsilon f_i}.
\end{aligned}
\label{eq:Smaxent}
\end{equation}
Here $\eta_i\in\{0,1\}$, $|\eta|=\sum_i\eta_i$, and $Z_Q(\epsilon)$ normalises the state. This is a member of the maximum-entropy
family at fixed particle number and prescribed mean occupations.
Indeed, maximising $-\sum_\eta p(\eta)\ln p(\eta)$ under these
constraints gives
$p(\eta)\propto\exp(\sum_i b_i\eta_i)$; Eq.~\eqref{eq:Smaxent} uses the particular biases $b_i=\epsilon f_i$. Under $\rho_*$, exchangeability and the fixed total particle number give
\begin{equation}
\operatorname{Cov}_{\rho_*}(n_i,n_j)
=\chi_L\left(\delta_{ij}-\frac{1}{L}\right),
\qquad
\chi_L=\nu(1-\nu)\frac{L}{L-1}.
\end{equation}
Consequently
\begin{equation}
\langle n_i\rangle_{\rho_f(\epsilon)}
=\nu+\epsilon \chi_L f_i+O(\epsilon^2),
\end{equation}
Expanding $\ln Z_Q$ in the biases, $\partial\langle n_i\rangle/\partial b_j|_{b=0}=\mathrm{Cov}_{\rho_*}(n_i,n_j)$, so $\langle n_i\rangle=\nu+\epsilon\sum_j\mathrm{Cov}_{\rho_*}(n_i,n_j)f_j+O(\epsilon^2)=\nu+\epsilon\chi_Lf_i+O(\epsilon^2)$, using $\sum_jf_j=0$.
Thus the circuit prepares the state in Eq.~\eqref{eq:Smaxent} exactly, while its density modulation is proportional to $f_i$
to leading order in $\epsilon$. Exact target densities can also be imposed at finite amplitude.
For $0<n_i^{\rm tar}<1$ and $\sum_i n_i^{\rm tar}=Q$, the required
biases minimise
\begin{equation}
\ln Z_Q(b)-\sum_i b_i n_i^{\rm tar},
\qquad
Z_Q(b)=\sum_{|\eta|=Q}\exp\!\left(\sum_i b_i\eta_i\right),
\end{equation}
over zero-sum bias vectors. The Hessian is the occupation covariance matrix, which is positive definite on this subspace. The biases therefore depend only on the target densities, not
on the disorder sample. Reflection-even targets give
reflection-even biases.

The second family consists of the affine states
\begin{equation}
\rho_f^{\rm aff}(\epsilon)
=\frac{P_Q(1+\epsilon A_f)P_Q}{d_Q},
\qquad
|\epsilon|\le\frac{1}{M_f},
\qquad
M_f=\max_{|\eta|=Q}|A_f(\eta)|.
\label{eq:Saff}
\end{equation}
These states are positive throughout the stated range. Their
densities are exactly $\nu+\epsilon\chi_Lf_i$, and their
slow-mode amplitude is exactly
$\epsilon\Tr(L_1^\dagger\rho_*A_f)$. Thus the residual has no
linear-response error at any allowed $\epsilon$. Since the
blind and reference cosine profiles have equal norms, their
initial root-mean-square density-distance ratio is exactly
$\kappa=|\epsilon_{\rm b}/\epsilon_{\rm r}|$. The restriction is
the available amplitude: for cosine profiles at fixed filling,
$M_f$ grows linearly with $L$, so this family requires
$\epsilon=O(L^{-1})$.

\paragraph{A deterministic circuit.}
Let $E_{i,q}$ be the sum of $\prod_{j\ge i}w_j^{\eta_j}$ over
occupations of sites $i,\ldots,L$ containing $q$ particles.
Separating the cases $\eta_i=0$ and $\eta_i=1$ gives
\begin{equation}
E_{i,q}=E_{i+1,q}+w_iE_{i+1,q-1},
\qquad
E_{L+1,q}=\delta_{q0},
\qquad
E_{1,Q}=Z_Q(\epsilon).
\end{equation}
We set $E_{i,q}=0$ for infeasible particle numbers. Conditional on having $q$ particles still to place, site $i$ is occupied with probability
\begin{equation}
p_{i,q}=\frac{w_iE_{i+1,q-1}}{E_{i,q}}. 
\end{equation}
Use $L$ site qubits and a counter of
$m=\lceil\log_2(Q+1)\rceil$ qubits, initialised in
$|0^L\rangle|Q\rangle$. At site $i$, apply the controlled rotation
\begin{equation}
V_i=\sum_q|q\rangle\langle q| \otimes R_y(\theta_{i,q})_i,
\qquad
\theta_{i,q}=2\arcsin\sqrt{p_{i,q}},    
\end{equation}
where
\begin{equation}
R_y(\theta)|0\rangle
=\cos\frac{\theta}{2}|0\rangle
 +\sin\frac{\theta}{2}|1\rangle.
\end{equation}
The formula for $\theta_{i,q}$ applies when $E_{i,q}>0$; angles on unreachable counter states can be chosen arbitrarily. Follow the rotation by the reversible decrement
\begin{equation}
W_i:
|\eta_i\rangle|q\rangle
\longmapsto
|\eta_i\rangle
|(q-\eta_i)\bmod 2^m\rangle.
\end{equation}
Along a branch with occupations $\eta$, let
$q_i=Q-\sum_{j<i}\eta_j$. Its squared amplitude telescopes
\begin{equation}
\prod_{i=1}^L
\frac{w_i^{\eta_i}E_{i+1,q_i-\eta_i}}{E_{i,q_i}}
=
\frac{\prod_iw_i^{\eta_i}}{Z_Q(\epsilon)}
E_{L+1,Q-|\eta|}.
\end{equation}
This vanishes unless $|\eta|=Q$. All amplitudes are
nonnegative because the rotation angles lie in $[0,\pi]$, and the counter ends in $|0\rangle$ on every branch. The circuit $W_LV_L\cdots W_1V_1$ therefore outputs
\begin{equation}
|\Psi_f\rangle|0\rangle,
\qquad
|\Psi_f\rangle
=\frac{1}{\sqrt{Z_Q(\epsilon)}}
 \sum_{|\eta|=Q}
 \prod_iw_i^{\eta_i/2}|\eta\rangle.
\end{equation}
Next apply $Z^{z_i}$ to each site, with independent fair bits $z_i$, and discard their record. Since $
2^{-L}\sum_{z\in\{0,1\}^L}
(-1)^{z\cdot(\eta-\xi)}
=\delta_{\eta\xi},$ this removes all coherences and yields $\rho_f(\epsilon)$ in Eq.~\eqref{eq:Smaxent}. Equivalently, copy each site onto an ancilla with a CNOT and trace out the ancillas. Computing the table $E_{i,q}$ costs $O(LQ)$ operations, and
the circuit has at most $O(LQ)$ conditional rotation angles.
If stage $j$ has operator-norm error $\varepsilon_j$, the
output error obeys
$\tfrac12\|\widetilde\rho-\rho_f(\epsilon)\|_1
\le\sum_j\varepsilon_j$.
Gate-level simulations reproduce $|\Psi_f\rangle$ to
$3\times10^{-16}$ for $(L,Q)$ up to $(9,5)$, using both
cosine and random biases.

\paragraph{Four qubits.}
For $L=4$, $Q=2$, and the blind profile
$f=(1,-1,-1,1)/\sqrt2$, set $w=\e^{\epsilon/\sqrt2}$. The coherent
state before phase averaging is
\begin{equation}
|\Psi_f\rangle=
\frac{
|1100\rangle+|1010\rangle+w|1001\rangle
+w^{-1}|0110\rangle+|0101\rangle+|0011\rangle
}{
\sqrt{w^2+w^{-2}+4}
}.
\end{equation}
Controls on the earlier site qubits replace the counter, as shown in Fig.~\ref{fig:circuit} in Appendix B.

Site 1 is occupied with probability
\begin{equation}
p_1=\frac{w^2+2}{w^2+w^{-2}+4},
\qquad
\theta_1=2\arcsin\sqrt{p_1}.
\end{equation}
The rotation on site 2 depends on the occupation of site 1
\begin{equation}
\theta_{2|0}
=2\arcsin\sqrt{\frac{w^{-2}+1}{w^{-2}+2}},
\qquad
\theta_{2|1}
=2\arcsin\frac{1}{\sqrt{w^2+2}}.
\end{equation}
Define
\begin{equation}
\theta_+=\frac{\theta_{2|0}+\theta_{2|1}}{2},
\qquad
\theta_-=\frac{\theta_{2|0}-\theta_{2|1}}{2}.
\end{equation}
The conditional rotation is implemented by
$R_y(\theta_+)$, a CNOT controlled by site 1,
$R_y(\theta_-)$, and a second CNOT. The net angle is
$\theta_{2|0}$ when site 1 is empty and $\theta_{2|1}$ when it is occupied.

Site 3 must be occupied after $00$ and empty after $11$. After $01$ or $10$, it requires the angle
\begin{equation}
\theta_3=2\arcsin\frac{1}{\sqrt{1+w^2}}.
\end{equation}
This is implemented by $R_y(\pi)$, two singly controlled $R_y(\theta_3-\pi)$ rotations, and a doubly controlled $R_y(\pi-2\theta_3)$ rotation. Their angles add to the required value on each branch. Finally, three CNOTs set the occupation of site 4 to the parity of the first three sites, completing the configuration with exactly two particles. Independent random $Z$ gates then produce the diagonal state. At $\epsilon=0.2$, the angles are
\begin{equation}
(\theta_1,\theta_{2|0},\theta_{2|1},\theta_3)
=(1.665214,1.848022,1.160546,1.429844),
\end{equation}
and the densities are
$(0.547139,0.452861,0.452861,0.547139)$.
The simulated circuit reproduces $|\Psi_f\rangle$ to
$2\times10^{-16}$ at $\epsilon=-1$, $0$, $0.2$, and $1$.

\paragraph{Conserving particle number throughout preparation.}
The rotation circuits end in the $Q$-particle sector but pass through other particle numbers. If $Q$ must be conserved at every step, the diagonal target state can instead be prepared
by classical sampling followed by loading. For the maximum-entropy family, sample a configuration
sequentially using the probabilities $p_{i,q}$. For the affine family, draw a uniform configuration with $Q$ particles and accept it with probability
\[
\frac{1+\epsilon A_f(\eta)}{1+|\epsilon|M_f}.
\]
The mean acceptance probability is
$1/(1+|\epsilon|M_f)\ge1/2$ throughout the positivity range in Eq.~\eqref{eq:Saff}. For either family, prepare $|1^Q0^{L-Q}\rangle$ and move
the particles to the sampled sites using SWAP gates.
These gates commute with total particle number. Pairing misplaced particles with empty target sites requires at most $\min(Q,L-Q)$ swaps between arbitrary sites, or $O(LQ)$ nearest-neighbor swaps. Averaging over the sampled configurations gives the desired diagonal state.

\paragraph{A heralded alternative.}
Another construction fixes the particle number by
postselection. Apply local rotations
$R_y(2\arcsin\sqrt{p_i})$, with
\begin{equation}
p_i=\frac{\zeta w_i}{1+\zeta w_i},
\qquad \zeta>0,
\end{equation}
and measure the total particle number. The outcome $Q$ heralds $|\Psi_f\rangle$ with probability
\begin{equation}
P_Q^{\rm succ}
=\frac{\zeta^QZ_Q(\epsilon)} {\prod_i(1+\zeta w_i)}.
\end{equation}
This probability is maximised when $\sum_i p_i=Q$. When the particle-number distribution admits a Gaussian approximation
\begin{equation}
P_Q^{\rm succ}\simeq
\left[2\pi\sum_i p_i(1-p_i)\right]^{-1/2}.
\end{equation}
For uniform weights, this becomes
$[2\pi L\nu(1-\nu)]^{-1/2}$.
The final phase averaging gives $\rho_f(\epsilon)$. This construction fixes $Q$ at its output but does not conserve particle number during the rotations.

\paragraph{Measuring the residual.}
Each profile is measured at $\pm\epsilon$: the symmetric difference removes the even-order nonlinear response, and the ratio of the two profiles removes the unknown decay and readout factors common to both.
Write $A_k=A_{f^{(k)}}$, with $k=1$ for the reference and $k=2$ for the blind profile. In the same disorder realisation, prepare each state at biases $\pm\epsilon$ and measure $A_1$
from the site-resolved densities. Define
\begin{equation}
R_{1k}(t)=
\frac{
\langle A_1\rangle_{k,+\epsilon}(t)
-\langle A_1\rangle_{k,-\epsilon}(t)
}{2\epsilon}.
\end{equation}
For the affine family, $[A_k,P_Q]=0$ gives
\[
\rho_k^{\rm aff}(\epsilon)
=\rho_\ast+\epsilon\rho_\ast A_k,
\qquad
\frac{\rho_k^{\rm aff}(\epsilon)
      -\rho_k^{\rm aff}(-\epsilon)}
     {2\epsilon}
=\rho_\ast A_k.
\]
This identity is exact throughout the positivity range. For the exponential family, the same symmetric difference equals $\rho_\ast A_k+O(\epsilon^2)$, because the even powers of $\epsilon$ cancel. Once the slow mode dominates, the response is
\begin{equation}
R_{1k}(t)\simeq
\Tr(A_1R_1)\,\e^{\lambda_1t}
\left[
\Tr(L_1^\dagger\rho_*A_k)+O(\epsilon^2)
\right],
\end{equation}
where the $O(\epsilon^2)$ term is absent for the affine family. Provided the reference overlap and the readout factor $\Tr(A_1R_1)$ are nonzero, the common factors cancel
\begin{equation}
\frac{R_{12}(t)}{R_{11}(t)}
\xrightarrow[t\to\infty]{}
\begin{cases}
r, & \text{affine preparations},\\
r+O(\epsilon^2), & \text{maximum-entropy preparations}.
\end{cases}
\label{eq:Sreadout}
\end{equation}
Thus the affine family measures $r$ at every allowed
amplitude, while the maximum-entropy family requires a small-amplitude extrapolation to remove the remaining bias. For the interacting lattice gas with $L=10$, $V=\gamma=4$, and fields in $[-4,4]$, the numerical response ratio agrees with $r$ within $10^{-6}$ once $t>10/(\lambda_1-\lambda_2)$, at the two modulation amplitudes tested.

\paragraph{Assumptions during preparation.}
The profiles, biases, and rotation angles depend only on the chosen target, $Q$, and the modulation amplitude, never on the fields, hoppings, or their distribution. During the coherent gate sequence, unwanted Hamiltonian evolution and dephasing must be negligible or controlled well enough to implement the prescribed gates. Switching off hopping alone is not sufficient in general, because the intermediate states need not be diagonal. Once the diagonal state has been prepared, diagonal fields, density-density interactions, and local dephasing leave it unchanged. The prescribed relaxation dynamics, including hopping, can then be applied.

\section{Numerical checks}
\label{sec:Schecks}

\begin{figure}[htb]
\centering
\includegraphics[width=0.8\textwidth]{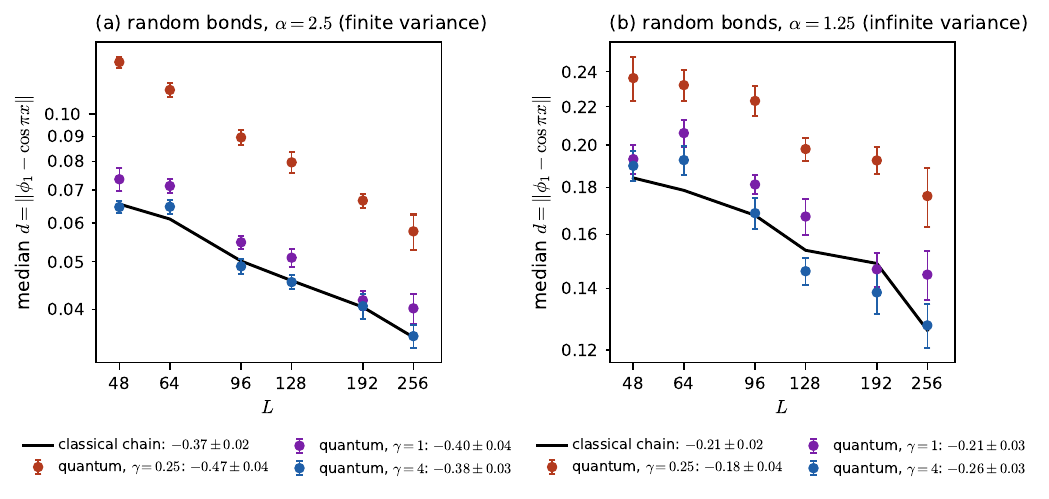}
\caption{Slow-mode distortion in the full quantum dynamics (points) and in the classical chain (line), random hoppings with (a) $\alpha=2.5$ and (b) $\alpha=1.25$. Legends give exponents over $L=48$-$256$.}
\label{fig:hopping}
\end{figure}

We test these predictions in two ways: with the exact quantum dynamics Eq.~\eqref{eq:closure}, and with the classical resistor chain, which reaches $L=2^{16}$. Table~\ref{tab:checks} compares the main predictions with the data. For random fields the quantum exponents match the resistor picture at all dephasing rates studied [Fig.~\ref{fig:bounded}]. For random hoppings over $L=48$-$256$ they agree with the classical chain within about two standard errors, except at $\alpha=2.5$, $\gamma=0.25$, where the quantum exponent $-0.47\pm0.04$ lies closer to the asymptotic $-1/2$ than the classical one $-0.37\pm0.02$ at these sizes [Fig.~\ref{fig:hopping}]. The N\'eel state and the imbalance used for this model in Ref.~\cite{Sgz} barely see the slow mode. For a smooth profile $f$, the alternating sum $\sum_i(-1)^if(i/L)$ is only a boundary term, so the N\'eel state's overlap with the slow mode and the slow mode's weight in the imbalance are each of order $1/L$. In a clean periodic chain of even length this overlap vanishes exactly by translation symmetry; for odd length it is of order $L^{-2}$~\cite{Sgz}. We therefore compare the preparations using the root-mean-square
density deviation defined in the main text. The $\cos(\pi x)$ component of a site-resolved density profile can monitor the late-time slow mode, but it vanishes initially for the $\cos(2\pi x)$ blind preparation and cannot by itself define the starting-distance ratio $\kappa$.

\begin{table}[htb]
\caption{Main predictions and their numerical tests. Exponents of the slow-mode distortion are fitted over the largest chains (up to $L=65536$). Quantum results use the exact Eq.~\eqref{eq:closure}.}
\label{tab:checks}
\begin{ruledtabular}
\begin{tabular}{lll}
quantity & prediction & result\\
\hline
exponent of $d$, $\alpha=4,\ 3,\ 2.5$ & $-0.5$ & $-0.50,\ -0.51,\ -0.49$ ($\pm0.01$)\\
exponent of $d$, $\alpha=2$ & $-0.448$ (with $\sqrt{\log L}$) & $-0.448\pm0.014$\\
exponent of $d$, $\alpha=1.67,\ 1.43,\ 1.25,\ 1.11$ & $-0.40,\ -0.30,\ -0.20,\ -0.10$ & $-0.39,\ -0.29,\ -0.15,\ -0.12$ ($\pm0.02$-$0.03$)\\
exponent of $d$, random fields, $h_0=1$-$16$ & $-0.5$ & $-0.495$ to $-0.498$\\
exponent of the rate & $-2$ & $-2.00$ to $-2.09$\\
exponent of $|r|$, quantum, $\gamma=1$ & $-0.5$ & $-0.49\pm0.02$\\
exponent of $|r|$, quantum, $\gamma=0.125$-$16$ & $-0.5$ & $-0.47$ to $-0.53$\\
exponent of the reach, random fields and $\alpha=1.25$ & $0.5$ and $0.2$ & $0.51$ and $0.21$\\
success probability, random fields & Eq.~\eqref{eq:Spwin}, no free parameter & agrees within statistical errors\\
residual distribution, $\alpha=1.25$ & symmetric stable, Eq.~\eqref{eq:Sstable} & Fig.~\ref{fig:heavy}(a)\\
success probability, $\alpha=1.25$, $256\le L\le4096$ & Eqs.~\eqref{eq:Sstable},~\eqref{eq:Szk} & within $0.022$\\
median reach, $\alpha=1.25$ & Eqs.~\eqref{eq:Sstable},~\eqref{eq:Szk} & $9\%$ at $L=256$, $3\%$ for $512\le L\le4096$\\
\end{tabular}
\end{ruledtabular}
\end{table}



\end{document}